\documentclass{article}

\usepackage{arxiv}

\usepackage[utf8]{inputenc} % allow utf-8 input
\usepackage[T1]{fontenc}    % use 8-bit T1 fonts
\usepackage{hyperref}       % hyperlinks
\usepackage{url}            % simple URL typesetting
\usepackage{booktabs}       % professional-quality tables
\usepackage{amsfonts}       % blackboard math symbols
\usepackage{nicefrac}       % compact symbols for 1/2, etc.
\usepackage{microtype}      % microtypography
\usepackage{lipsum}
\usepackage{graphicx}
\graphicspath{ {./images/} }

\title{Autoregressive Images Watermarking through Lexical Biasing: An Approach Resistant to Regeneration Attack}

\author{
 Siqi Hui \\
  Xi'an Jiaotong University\\
  \texttt{huisiqi@stu.xjtu.edu.cn} \\
   \And
   Yiren Song \\
   National University of Singapore \\
   \texttt{songyiren725@gmail.com} \\
   \And
   Sanping Zhou\\
  Xi'an Jiaotong University\\
   \texttt{spzhou@xjtu.edu.cn}\\
    \And
   Ye Deng\\
  Southwestern University of Finance and Economics\\
   \texttt{dengye@stu.xjtu.edu.cn}\\
      \And
   Wenli Huang\\
Ningbo University of Technology\\
   \texttt{huangwenwenlili@126.com}\\
      \And
   Jinjun Wang\\
Xi'an Jiaotong University\\
   \texttt{jinjun@mail.xjtu.edu.cn}\\
}
\usepackage[linesnumbered,ruled]{algorithm2e}
\usepackage{multirow}
\usepackage{amsmath}
\newcommand{\encoder}{\mathcal{E}}

\begin{document}
\maketitle

\begin{abstract}
Autoregressive (AR) image generation models have gained increasing attention for their breakthroughs in synthesis quality, highlighting the need for robust watermarking to prevent misuse. However, existing in-generation watermarking techniques are primarily designed for diffusion models, where watermarks are embedded within diffusion latent states. This design poses significant challenges for direct adaptation to AR models, which generate images sequentially through token prediction. Moreover, diffusion-based regeneration attacks can effectively erase such watermarks by perturbing diffusion latent states. To address these challenges, we propose \textbf{L}exical \textbf{B}ias \textbf{W}atermarking (LBW), a novel framework designed for AR models that resists regeneration attacks. LBW embeds watermarks directly into token maps by biasing token selection toward a predefined \textit{green list} during generation. This approach ensures seamless integration with existing AR models and extends naturally to post-hoc watermarking. To increase the security against white-box attacks, instead of using a single green list, the green list for each image is randomly sampled from a pool of green lists. Watermark detection is performed via quantization and statistical analysis of the token distribution. Extensive experiments demonstrate that LBW achieves superior watermark robustness, particularly in resisting regeneration attacks.
\end{abstract}
\section{Introduction}
\label{sec:intro}
Recent diffusion models have demonstrated remarkable success across a wide range of generative tasks, including text-to-image synthesis\cite{sdxl, sd3, ldm}, controllable generation\cite{zhang2024ssr, zhang2024fast, zhang2025easycontrol},  image editing\cite{yang2024editworld, huang2025photodoodle,shi2024seededit,kawar2023imagic, stablemakeup, zhang2025stable}, and video generation\cite{zhang2024show, song2025makeanything, song2024processpainter, wan2024grid}. While diffusion models~\cite{flux2024,betker2023improving} have dominated the landscape, autoregressive (AR)-based frameworks have emerged as a compelling alternative, achieving state-of-the-art image quality~\cite{tian2025var,han2024infinity,yu2024rar}. Moreover, AR-based image modeling can be seamlessly integrated with AR-based language modeling frameworks, enabling powerful multimodal applications~\cite{xie2024showo,zhou2024transfusionpredicttokendiffuse}. However, their ability to generate highly realistic images concerns potential misuse, including deep-fakes and misinformation. To ensure traceability and prevent abuse, it is crucial to develop effective watermarking techniques for images generated by AR models.

Existing watermarking techniques can be categorized into post-hoc and in-generation approaches. Post-hoc watermarking embeds watermarks into pre-generated images via imperceptible perturbations~\cite{Rahman_2013, zhang2019rivagan, fernandez2022watermarking}, whereas in-generation watermarking integrates watermarks directly into the diffusion-based image synthesis process by modifying intermediate states~\cite{wen2023treering, ci2024ringid, huang2024robinrobustinvisiblewatermarks}. While effective in diffusion models, in-generation methods are incompatible with AR models, which generate images sequentially via token prediction rather than refining continuous latent representations. Consequently, watermarking within AR-based image generation remains an open challenge. Moreover, regeneration attacks pose a significant threat to diffusion-based in-generation watermarking by disrupting their latent representations where watermarks are embedded~\cite{nie2022diffusionmodelsadversarialpurification, liu2024ctrlregen}. In contrast, AR models, which generate images through discrete token prediction, offer a fundamentally different mechanism that may enhance robustness against such attacks. This motivates the development of a watermarking method specifically tailored for AR models.

The primary challenge in embedding watermarks in AR-generated images is determining where to introduce the watermark so that it remains detectable. A key observation is that AR models quantize images into token maps, and when an AR-generated image is re-encoded, a significant portion of the original tokens can be recovered (see Fig~\ref{fig:observation2}). This suggests that watermark information can be embedded in the token map, and subsequently detected by re-quantizing the watermarked image and analyzing the token distribution. Additionally, we observe that minor perturbations within a controlled range on the token map do not significantly degrade image quality (see Fig~\ref{fig:observation1}, ~\ref{fig:observation1_images}). These observations motivate us to embed watermarks in token maps of AR models. 

In this paper, drawing inspiration from text watermark techniques, we propose a novel framework called \textbf{L}exical \textbf{B}ias \textbf{W}atermarking (LBW), which embeds watermarks in the token map by introducing a controlled bias in token selection during the autoregressive prediction process. Specifically, we partition the token vocabulary into red and green lists and encourage the model to favor green tokens during prediction by applying a \textbf{soft token biasing strategy}, which increases the logits of green tokens with a constant to enhance their likelihood of being sampled. Instead of utilizing dynamic green list~\cite{kirchenbauer2023reliability,hu2023unbiased}, we adopt a \textbf{global token partition strategy}, which maintains a predetermined green list throughout the entire token generation process. This design ensures compatibility with random token prediction processes~\cite{pang2024randardecoderonlyautoregressivevisual,yu2024rar} and enhances robustness against global image watermark removal attacks. Furthermore, our method naturally extends to post-hoc watermarking by leveraging the VQ-VAE-based image reconstruction process. After an image is quantized into a token map, we embed the watermark by replacing red tokens with their nearest green counterparts. The modified token map is then used to reconstruct the image, effectively embedding the watermark in a post-hoc manner. 

To further enhance resistance against white-box attacks, we introduce a \textbf{multi-green-list} strategy rather than relying on a single green list. During watermarking, one green list is randomly selected from multiple green lists, and these green lists are predefined such that each token has an equal probability of being a green token across the entire green list pool. Empirical results in Figure~\ref{fig:token_frequency} confirm that our multi-list strategy produces token distributions nearly indistinguishable from those of clean images, whereas the single-list strategy exhibits detectable biases that can be easily inferred by an adversary. 

For watermark detection, we apply a \textbf{z-score hypothesis test} to evaluate the proportion of green tokens in the token map quantized from watermarked images. Specifically, we evaluate each green list in the pool by computing the proportion of its tokens present in the token map. Given the high token consistency in AR-generated images (see Fig~\ref{fig:observation2}), any statistically significant deviation from the expected green token ratio provides strong evidence of watermark presence. This detection method is lightweight, requiring only VQ-VAE without access to transformer-based generative models, making it suitable for both in-generation and post-hoc watermarking. Experimental results demonstrate that our approach achieves state-of-the-art robustness against both conventional and regeneration watermark removal attacks, particularly CtrlRegen~\cite{liu2024ctrlregen}, a strong attack designed to erase watermarks embedded in diffusion models. This highlights the effectiveness of our method in providing resilient watermarking for AR-generated images.

In this work, we make the following key contributions:
\begin{itemize}
    \item  To the best of our knowledge, this is the first study to explore watermarking for the AR image generation process, ensuring seamless integration without disrupting the iterative token prediction mechanism.
    \item We propose LBX, a unified framework that introduces lexical bias in AR-based image generation and reconstruction processes. We also introduce a multi-green-list strategy to increase the security against white-box attacks.
    \item  Extensive experiments demonstrate that our method achieves comparable robustness to baseline watermarking techniques against conventional attacks while exhibiting superior resilience against regeneration attacks.
\end{itemize}

\section{Related Works}
\noindent\textbf{Image watermarking methods.}

Image watermarking ensures digital content authenticity and security, typically categorized as post-hoc or in-generation watermarking. Post-hoc methods embed watermarks into pre-generated images via pixel-based (e.g., LSB~\cite{LSB}) or frequency-based techniques (e.g., DwtDct and DwtDctSvd~\cite{dwtdctsvd}), with recent approaches leveraging deep learning (e.g., RivaGAN~\cite{zhang2019rivagan}, SSL~\cite{fernandez2022watermarkingimagesselfsupervisedlatent}, StegaStamp~\cite{Tancik2020stegastamp}). In-generation methods integrate watermarks into the image synthesis process by modifying intermediate states, particularly in diffusion models\cite{wen2023treering,ci2024ringid,huang2024robinrobustinvisiblewatermarks} or VAE decoders\cite{Fernandez_2023_stablesignature,ci2024wmadapter}. However, these methods are incompatible with AR models, which generate images via sequential token prediction. Besides, they are vulnerable to diffusion-based regeneration attacks. We firstly explores in-generation watermarking for AR image generation, demonstrating superior robustness against regeneration attacks.

\noindent\textbf{Text watermark methods for LLMs.}
Watermarking LLMs typically involves modifying logits or token sampling to embed watermarks within the generated text. KGW\cite{kirchenbauer2023watermark} partitions the vocabulary into "green" and "red" lists using a hash-based selection strategy, biasing token selection toward green-listed tokens. EWD\cite{lu2024entropy} enhances detection by assigning higher weights to low-entropy tokens. To minimize text distortion, SWEET\cite{liu2024adaptive} and Adaptive Watermark\cite{lee2024wrotecodewatermarkingcode} avoid embedding watermarks in low-entropy positions, while BOW~\cite{wouters2023optimizing} selectively skips red tokens with high probabilities. WinMax\cite{kirchenbauer2023reliability} applies a sliding window approach to defend against text mixing attacks, while semantic grouping techniques\cite{liu2024semanticinvariantrobustwatermark,he2024can} cluster similar tokens to resist semantic-invariant modifications. Unlike these adaptive methods, our approach utilizes a globally fixed green list, ensuring compatibility with AR models, which generate tokens in any order. This design enhances robustness against global image watermark removal attacks, including blurring and DiffPure\cite{nie2022diffusionmodelsadversarialpurification}.

\noindent\textbf{Autoregressive visual models.}
Early research on autoregressive (AR) image generation\cite{chen2020generative,gregor2014deep,parmar2018image} modeled 2D images as 1D pixel sequences, generating pixels in a row-wise raster scan order. Recent advancements leverage VQ-VAE-based tokenization\cite{oord2018neuraldiscreterepresentationlearning}, where models like VQGAN\cite{esser2021taming} employ decoder-only transformers to predict sequences of discrete latent tokens. Similar paradigms include VQVAE-2\cite{razavi2019generating} and RQ-Transformer~\cite{lee2022rqvae}. To overcome the limitations of unidirectional raster-scan generation, VAR\cite{tian2025var} introduced a multi-scale residual token map, improving spatial coherence. Further refinements, such as RAR\cite{yu2024rar} and RandAR~\cite{pang2024randardecoderonlyautoregressivevisual}, shuffle token generation orders during training, facilitating bidirectional token dependencies and enhancing contextual coherence. Our proposed LBW could be seamlessly integrated with AR models that employ both single-scale and multi-scale token maps as well as predefined and randomized token generation orders.

\section{Method}
In this section, we provide a detailed explanation of LBW, which embeds watermarks into token maps through both in-generation and post-hoc approaches, along with the corresponding watermark detection process. Section~\ref{sec:prelimi} provides an overview of the image quantization and token prediction processes fundamental to AR-based image synthesis. Section~\ref{sec:observation} presents two key properties of AR models that enable robust watermark embedding while preserving image quality. Section~\ref{sec:method} details our in-generation and post-hoc watermarking methods, along with their detection process.

\subsection{Preliminary}
\label{sec:prelimi}
\noindent\textbf{Tokenization.}
Current AR image models~\cite{esser2021taming,tian2025var,yu2024rar} leverage VQ-VAE~\cite{van2017vqvae} to represent continuous images $x\in\mathbb{R}^{H\times W\times 3}$ as discrete tokens $(x_1,x_2,...,x_T)$ in latent space, where each $x_i \in [V]$ is an integer from a vocabulary of size $V$. Given an input image $x$, it is first encoded in a feature map $f\in \mathbb{R}^{h\times w \times C}=\encoder(x)$, where $\encoder(\cdot)$ is the encoder. The quantizer $\mathcal{Q}(\cdot)$ with a learnable codebook $Z\in \mathbb{R}^{V\times C}$ then maps the feature map to a discrete token map $q\in [V]^{h\times w}$ or a stack of token maps $\{ q_k \}_{k=1}^{K}\mid q_k \in [V]^{h_k \times w_k}$, based on the single-scale or multi-scale quantization process they introduce. The \textbf{single-scale quantization process $q=\mathcal{Q}(f)$} maps each feature vector $f^{(i,j)}$ to a code index $q^{(i,j)}$ by finding the nearest code in the codebook $Z$ using Euclidean distance:
\begin{equation}
    q^{\text{(}i,j\text{)}}=\underset{v\in [ V ]}{arg\min}||\text{lookup}( Z,v ) -f^{\text{(}i,j\text{)}}||_2,
\end{equation}
where lookup($Z,v$) fetches the $v$-th vector from the codebook $Z$. This process produces the approximated feature map $\hat{f}$, which is decoded by $\mathcal{D}(\cdot)$ to generate the reconstructed image $\hat{x}$:
\begin{equation}
\begin{aligned}
        \hat{f}=\text{lookup}(Z,q), \quad \hat{x}=\mathcal{D}(\hat{f}). 
\end{aligned}
\end{equation}
The \textbf{multi-scale quantization process $\{ q_k \}_{k=1}^{K}=\mathcal{Q}_K(f)$} progressively derives token maps at each scale. A set of scale parameters ${(h_k,w_k)}_{k=1}^{K}$ defines the map resolutions in ascending order, with $h_K=h$ and $w_K=w$ representing the largest scale. The token map $q_k$ for scale $k$ is computed by quantizing the residual feature map $r_k$, which is derived by subtracting the aggregated sum of the reconstructed residual feature maps from preceding scales (each upscaled to the maximum resolution) from the original feature map:
\begin{equation}
\begin{aligned}
    r_k=&f-\sum_{i=1}^{k-1}{\text{interpolate}(\hat{r_i},h_K,w_K )}, \\
    \hat{r_k} =& \text{lookup}(Z,q_k),
\end{aligned}
\end{equation} 
where $\hat{r_k}$ denotes the approximated residual feature map at scale $k$. Finally, the image is reconstructed by decoding $\hat{f}$: $\hat{x} = \mathcal{D}(\hat{f})$. 

\noindent\textbf{Token prediction.}
AR models synthesize images through sequential token prediction after modeling images as sequences of discrete latent tokens. They utilize transformers to model the conditional probability distribution of token generation, formulated as:
\begin{equation}
\begin{aligned}
    p_{\theta}( x_t|\mathbf{X}_t ) = \text{softmax} ( l_{\theta} (\mathbf{X}_t) ),
\end{aligned}
\end{equation}
where $l_{\theta} (\mathbf{X}_t)~\in~\mathbb{R}^V$ is the generated logit at step $t$, and $\mathbf{X}_t\subseteq\{x_1,...,x_{t-1}\}$ is the subset of previous generated tokens. The token sequence is generated through iterative sampling from the conditional probability. For simplicity, we use $l_t$ to denote the logits $l_\theta(\mathbf{X}_t)$ in the rest of the paper. For \textbf{single-scale token prediction}, tokens in $\mathbf{X}_t$ are generated either in a predefined order (e.g., raster-scan) or in a randomly permuted sequence. \textbf{Multi-scale token prediction} conditions token generation on all tokens from preceding scales, while allowing tokens within the same scale to be generated in parallel. For a more comprehensive discussion on multi-scale quantization and token prediction, please refer to~\cite{tian2025var} and~\cite{lee2022rqvae}.

\subsection{Observations} \label{sec:observation} To investigate the behavior of VQ-VAE and evaluate its potential for embedding robust watermarks, we conducted two primary analyses. \textbf{Observation 1} examines the consistency between token maps obtained by quantizing the original images and those derived from their reconstructed counterparts, providing insights into the feasibility of watermark embedding and detection in AR visual models. \textbf{Observation 2} investigates the VQ-VAE codebook by analyzing the impact of vocabulary size reduction on image reconstruction quality. This analysis determines whether a compact codebook can facilitate watermarking while preserving image fidelity.

Experiments were performed on three AR models featuring different vocabulary sizes: VQ-GAN~\cite{esser2021taming} (13,678 tokens), VAR~\cite{tian2025var} (4,096 tokens), and RAR~\cite{yu2024rar} (1,024 tokens). The evaluation dataset comprises 5,000 images randomly sampled from 100 ImageNet classes, with 50 images per class.

\begin{figure}[tbp] 
\centering 
\includegraphics[width=0.55 \linewidth]{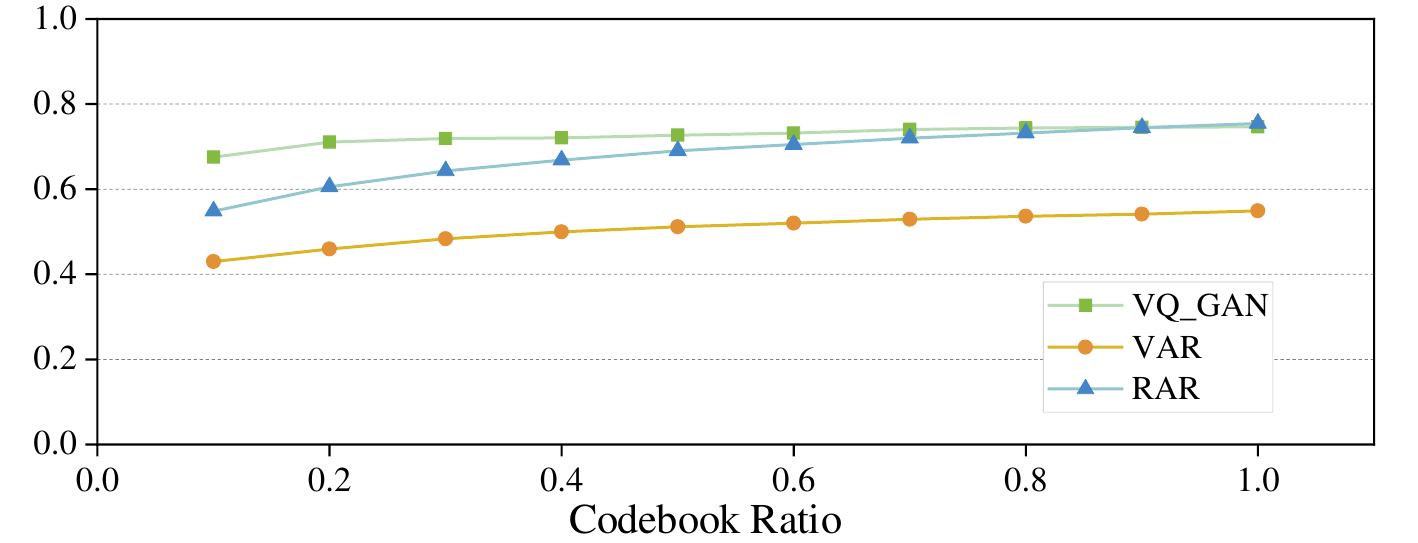} \caption{\textbf{Observation 1:} Token consistency for VQ-GAN, VAR, and RAR across various codebook ratios ranging from 0.1 to 1.0.} 
\label{fig:observation2} 
\end{figure}
\begin{figure}[tbp] 
\centering 
\includegraphics[width=1.0\linewidth]{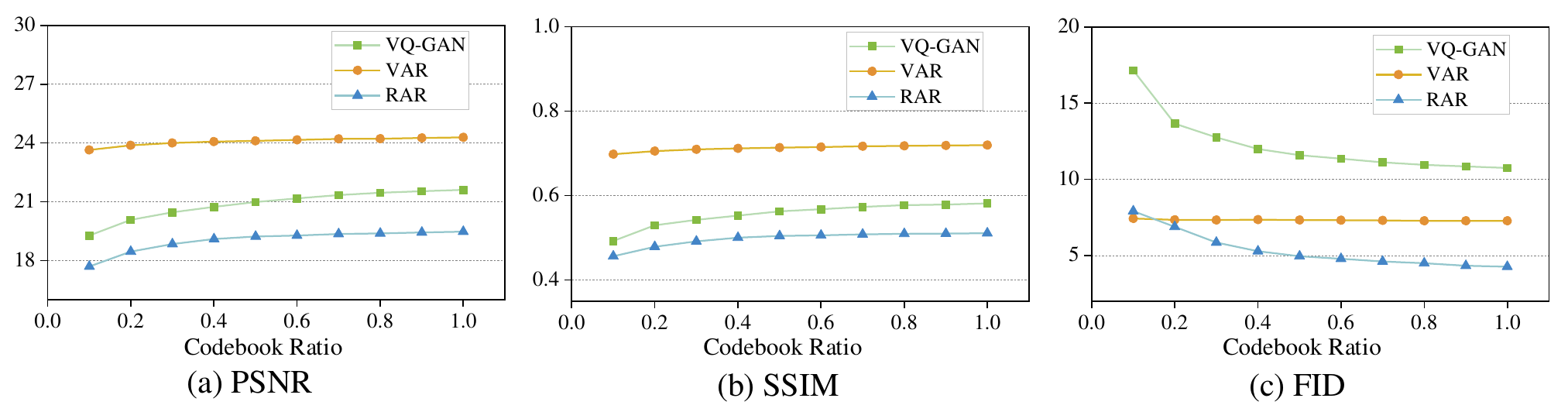} \caption{\textbf{Observation 2:} Image quality metrics (PSNR, SSIM, and FID) for AR reconstructed images across various codebook sizes, ranging from 0.1 to 1.0.} \label{fig:observation1} 
\end{figure}

\noindent\textbf{Observation 1: Token consistency.}
To evaluate the potential for watermark embedding and detection, we reconstructed images from the evaluation dataset and compared the token maps produced by quantizing both the input image 
$x$ and its reconstruction $\hat{x}$, yielding token maps $q$ and $\hat{q}$, respectively. The token consistency is defined as the proportion of matching tokens between these token maps. Notably, this consistency was assessed while progressively reducing the vocabulary size during the reconstruction process. For VAR, which employs a multi-scale quantization process, token consistency was evaluated at its largest scale. As illustrated in Fig~\ref{fig:observation2}, token consistency remains robust across various codebook sizes. However, RAR exhibits a more pronounced decline in consistency as the vocabulary decreases, likely due to its smaller codebook (1024) being more susceptible to quantization errors. These findings indicate that AR models employing VQ-VAEs maintain sufficient token consistency to preserve a substantial portion of the original token sequence even under significant vocabulary reduction. Consequently, watermark information can be reliably embedded into token maps and subsequently detected by quantizing images into these maps again.

\noindent\textbf{Observation 2: Codebook redundancy.}
\begin{figure}[tp] 
\centering 
\includegraphics[width=1.0\linewidth]{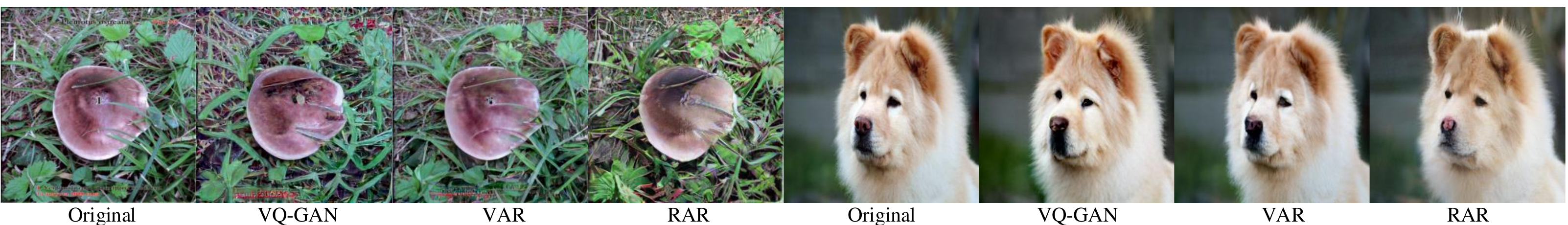} \caption{Reconstructed images using only 10\% of the original codebook size.} \label{fig:observation1_images}
\end{figure}
To further assess the influence of codebook reduction on image quality, we decreased the codebook ratio during image reconstruction and evaluated PSNR, SSIM, and FID metrics of reconstructed images. As shown in Fig~\ref{fig:observation1}, a moderate degradation in image quality is observed as the codebook size decreases, and the overall performance remains stable. Notably, even when the vocabulary of VAR was reduced to just 10\%, the reconstructed images exhibited minimal quality loss relative to those produced using the full codebook. This finding indicates that a compact codebook is viable for watermarking applications without substantially compromising image quality. Additionally, Fig~\ref{fig:observation1_images} presents the reconstructed images obtained using only 10\% of the vocabulary, further demonstrating the feasibility of this approach.

\subsection{Watermarking through lexical biasing}
\label{sec:method}
Building on our previous analysis, we observe that tokens used for image generation can be recovered by encoding and quantizing AR-generated images into token maps again (see Fig~\ref{fig:observation2}). This finding suggests that if watermark information is embedded in the token map, it should be preserved and detected through this re-quantization process. In this paper, drawing inspiration from text watermarking techniques, we aim to embed the watermark information through biasing AR models to use specific tokens during the autoregressive token prediction process. 

% The image quality could be controlled by adjusting the degree of token selection bias (see Figure~\ref{fig:observation1}). Then, the presence of a watermark can then be verified by assessing the statistical significance of the existence of these tokens in re-quantized token maps. 

\noindent\textbf{A direct approach.} 
Formally, the codebook of AR models can be partitioned into a green list \( G \) and a red list \( R \). We aim to bias the model toward selecting tokens from \( G \) during AR image synthesis. A simple yet effective approach is to mask the logits of red list tokens while retaining those of the green list.  
Specifically, at each timestep \( t \), the model predicts logits \( l_t \) based on previously generated tokens \( \mathbf{X}_t \), which are then converted into a discrete probability distribution for sampling the next token \( x_t \). To enforce token selection from \( G \), the logits of red list tokens are set to negative infinity, modifying the logit vector \( \hat{l}_t \) as follows:  
\begin{equation}
\hat{l}_{t(i)} =
\begin{cases}
    -\infty, & i \in R \\
    l_{t(i)}, & i \in G.
\end{cases}
\end{equation}  
\noindent This modification ensures that the softmax function assigns zero probability to tokens in \( R \), thereby restricting the model to sample \( x_t \) exclusively from the green list \( G \). 
Unlike text watermarking, where the green and red lists dynamically adjust based on a hash of previously generated tokens, we employ a \textbf{global token partitioning strategy}, in which the same green and red lists are uniformly applied to all tokens. This approach is driven by three considerations. \textbf{1)~Re-quantization Loss:} The re-quantization process introduces token variations. When a previously generated token changes, the corresponding green list for subsequent tokens would also shift, which hampers the robustness. \textbf{2)~Vulnerability to image watermarking attacks:} Unlike local text removal attacks, image watermarking attacks—such as blurring or CtrlRegen~\cite{liu2024ctrlregen}—impact the entire image, leading to substantial token variations across the token map, which also weakens the robustness of token-dependent hashing schemes. \textbf{3)~Compatibility with AR models:} Certain AR models generate tokens in a non-sequential, randomized order~\cite{pang2024randardecoderonlyautoregressivevisual,yu2024rar} rather than following a strictly predefined sequence. In such cases, if the token generation order is unknown, dynamically determining the green lists becomes impractical.

\noindent\textbf{Watermark detection.} 
Watermark presence is verified by statistically assessing the occurrence of specific tokens in the quantized token map, requiring only VQ-VAE and the predefined green list 
$G$ without access to transformer models. We formulate the verification process as a hypothesis-testing problem. We define the null hypothesis: $H_0: \ The \ image \ was \ generated \ without \ any \ bias \ toward \ the \ green \ list$. To evaluate this hypothesis, we perform a one-proportion z-test on the number of green tokens in the quantized token map. Let \( \gamma \) represent the proportion of green tokens used for watermarking. Under $H_0$, the expected number of green tokens in a token map $q\in \mathbb{R}^{h\times w}$ follows a binomial distribution with mean \( \gamma \cdot h \cdot w \) and variance \( \gamma (1 - \gamma) \cdot h \cdot w \). Denoting the observed number of green tokens in the token map as \( |s|_G \), the z-score for watermark detection is computed as:
\begin{equation}
    z = \frac{|s|_G - \gamma \cdot h \cdot w}{\sqrt{\gamma \cdot (1 - \gamma) \cdot h \cdot w}}.
\label{eq:zscore}
\end{equation}
\noindent By setting a threshold $z_{th}$, we reject $H_0$ and confirm watermark presence if $z>z_{th}$. 

% For instance, consider an image quantized into a \( 16 \times 16 \) token map, generated with a 50\% green token ratio (\(\gamma = 0.5\)). If the detection threshold is set to \( z_{\text{th}} = 4 \), our method can successfully identify watermarks as long as the token consistency between the generated and re-quantized token maps larger than 0.25, which could be satisfied by all AR models we have used (see Figure~\ref{fig:observation2}). 

\noindent\textbf{Enhancing watermark via soft biasing.} 
The direct approach enforces the exclusive use of green-listed tokens during token prediction. However, when the green list is overly constrained, the limited token vocabulary reduces the expressive capacity of AR models, leading to declined image quality or even generation failures (Fig~\ref{fig:simple_vs_lbw}). Moreover, restricting token selection disrupts the natural token distribution expected by VQ-VAE, thereby reducing token consistency, which in turn compromises watermark detectability and robustness (Fig~\ref{fig:token_consistency}).
To address these limitations, we employ a \textbf{soft token biasing} strategy that encourages the selection of green tokens without completely excluding red tokens. After predicting logits, instead of forcing red token logits to negative infinity, we introduce a bias constant \( \sigma \) to increase the logits of green tokens as:
\begin{equation}
\hat{l}_{t( i )}=
\begin{cases} 
    l_{t(i)} ,\quad & i \in R \\
    l_{t( i )} + \sigma,\quad & i \in G.
\end{cases}
\end{equation}
\noindent This ensures that when the transformer model exhibits high logits on the red list (high urge of using a red token), the added bias minimally influences token selection, preserving the natural AR generation process. As a result, this method maintains image quality and results in better token consistency and detectability. We also compute the z-score of the total number of green tokens in the re-encoded token map and compare it against a predefined threshold to determine the presence of a watermark.

\noindent\textbf{Post-hoc watermarking via token substitution.}
Our approach could be extended to support post-hoc watermarking for existing images by simply substituting quantized red tokens with green tokens. Formally, given an input image \( x \), we first quantize it into a token map \( q \). The watermark is then embedded by replacing each red token \( q_{(i)} \) with its nearest green token, determined by the Euclidean distance in the token embedding space, ensuring minimal distortion:
\begin{equation}
    q'_{(i)} = \underset{g \in G}{\arg\min} \| \text{lookup}(Z, g) - \text{lookup}(Z, q_{(i)}) \|_2.
\end{equation}
% \noindent The detection process remains unchanged, following the same z-score-based method as our in-generation watermarking.

\begin{algorithm}[tbp]\scriptsize
\caption{Generate Green List Matrix \( M \)}
\label{alg:generate_M}
\KwIn{Number of green lists \( N \), green token ratio \( \gamma \), vocabulary size \( V \)}
\KwOut{Binary matrix \( M \in \{0,1\}^{N \times V} \)}
% Step 1: Initialization
Randomly initialize matrix \( M \) such that each row \( i \) satisfies \(\sum_{j=1} M_{ij} = \gamma V\)\;
Set threshold \(\theta \leftarrow \gamma N\)\;
\Repeat{\text{convergence or maximum iterations reached}}{
    \For{\( i \leftarrow 1 \) \KwTo \( N \)}{
        Compute token frequency vector $f$: \(f[j] \leftarrow \sum_{i=1} M_{ij}\)\;
        \tcp{Identify indices for tokens with excessively high frequency}
        $\text{one\_to\_zero} \leftarrow \{ j \mid \text{token\_frequency}[j] > \theta \text{ and } M_{ij} = 1 \}$\;
        \tcp{Identify indices for tokens with too low frequency}
        $\text{zero\_to\_one} \leftarrow \{ j \mid \text{token\_frequency}[j] < \theta \text{ and } M_{ij} = 0 \}$\;

        \(K \leftarrow \min(|\text{one\_to\_zero}|, |\text{zero\_to\_one}|)\)\;

        \For{\( k \leftarrow 1 \) \KwTo \( K \)}{
            Set \( M_{i, \text{zero\_to\_one}[k]} \leftarrow 1 \)\;
            
            Set \( M_{i, \text{one\_to\_zero}[k]} \leftarrow 0 \)\;
        }
    }
}
\Return \( M \)
\end{algorithm}

\noindent\textbf{Multiple green lists.}
To prevent our method against white-box attacks, we propose a multiple green list strategy. Concretely, a set of $N$ green lists $\{G_i\}_{i=1}^N$ is established, from which one green list is randomly selected for watermark encoding. This set of green lists can be represented as a binary matrix $M\in\{0,1\}^{N\times V}$ as 
\begin{align}
M_{ij} &= 
\begin{cases}
1, \quad \text{if } j \in G_i, \\
0, \quad \text{otherwise},
\end{cases}
\quad \forall i,\forall j, \quad \text{s.t.}
\sum_{j=1}^V M_{ij} = \gamma V, \ \forall i, \  \text{and } 
\sum_{i=1}^N M_{ij} = \gamma N, \ \forall j.
\end{align}
Each row of $M$ corresponds to a green list pool, with the first constraint ensuring that each green list maintains a consistent green token ratio $\gamma$, while the second guarantees that each token is selected as a green token with equal probability across the green lists. This design aligns the token distribution of watermarked images with that of clean images, thereby reducing the risk of reverse-engineering the watermark. However, finding such a matrix exactly satisfying the above constraints requires solving the 0-1 integer programming problem, which is generally NP-hard and may have no feasible solution. To efficiently generate a matrix that approximately meets these constraints, we employ the following algorithm:

In our experiments, we use $N=32$ green lists and find it suffices to defend green list estimation attacks~\ref{fig:token_frequency}. During detection, the green token ratio is computed with respect to each green list in the pool, and the maximum observed ratio is used to calculate a z-score to justify the presence of the watermark. Notably, the detection requires only a convolutional image encoder to extract token maps, enabling efficient detection even when multiple green lists are employed.

\section{Experiments}

\subsection{Experimental Setting}
We evaluate our watermarking methods using VQ-GAN, RAR, and VAR. For VAR, watermarks are embedded in the largest-scale token map. We apply three watermarking variants: \textbf{LBW-Hard} (strict green token enforcement), \textbf{LBW-Soft} (soft bias toward green tokens), and \textbf{LBW-Post} (post-hoc token substitution). Green token ratios $\gamma$ are set to 0.1 for LBW-Post and LBW-Soft on VAR and RAR, and 0.2 for LBW-Hard on VQ-GAN. LBW-Soft uses bias constants $\sigma$ of 7, 4, and 8 for VAR, VQ-GAN, and RAR, respectively. To further improve the generation
quality, LBX-Soft adopts a bias constant $\sigma$ of 7, 4, and 8 for
VAR, VQ-GAN, and RAR. We use $N=32$ green lists for embedding and detection, with consistent green list configurations across in-generation and post-hoc methods.   
We compare against state-of-the-art watermarking methods: DwtDct~\cite{al2007combined}, DwtDctSvd~\cite{dwtdctsvd}, RivaGAN~\cite{zhang2019rivagan}, SSL~\cite{fernandez2022watermarkingimagesselfsupervisedlatent}, Tree-Ring~\cite{wen2023treering}, and WatermarkDM~\cite{zhao2023recipe}. Conventional attacks include Gaussian noise and blur, ColorJitter, geometric transformations (crop, resize, rotation), and JPEG compression. Regeneration attacks comprise VAE reconstruction (Stable Diffusion 1.5), DiffPure~\cite{zhou2024transfusionpredicttokendiffuse}, and CtrlRegen~\cite{liu2024ctrlregen}. Detailed attack settings can be found in the appendix~\ref{sec:attacks}. 
We evaluate performance on the ImageNet dataset. For post-hoc watermarking, 1,000 images (10 per 100 classes) are watermarked. For in-generation methods, 1,000 watermarked and 1,000 clean images are generated conditionally on class labels aligned with post-hoc experiments. Detection is evaluated by ROC-AUC and TPR@1\%FPR, averaged over five runs with different seeds to ensure robustness and reproducibility.

\subsection{Main Results}
\begin{table*}[htbp]\scriptsize
\centering
\caption{Comparative evaluation of watermarking methods under conventional and regeneration attacks. The table presents AUC and TPR@1FPR metrics, where TPR@1FPR is abbreviated as T@1F. Best results are \textbf{bolded}. Our proposed method exhibits superior robustness, particularly against regeneration attacks.}
\begin{tabular}{cc|c|c|ccccc|cccc} 
\hline
\multicolumn{2}{c|}{\multirow{2}{*}{Method}}                                                                                      & \multirow{2}{*}{Metric} & \multirow{2}{*}{Clean} & \multicolumn{5}{c|}{Conventional Attack}                                           & \multicolumn{4}{c}{Regeneration Attack}                           \\ \cline{5-13} 
\multicolumn{2}{c|}{}                                                                                                             &                         &                        & Gaus           & Color          & Geo            & JPEG           & AVG            & VAE            & Diff           & Ctrl           & AVG            \\ \hline
\multicolumn{2}{c|}{\multirow{2}{*}{dwtDct}}                                                                                      & AUC                     & 0.978                  & 0.906          & 0.333          & 0.631          & 0.520          & 0.598          & 0.521          & 0.485          & 0.496          & 0.501          \\
\multicolumn{2}{c|}{}                                                                                                             & T@1F                    & 0.920                  & 0.714          & 0.091          & 0.055          & 0.005          & 0.216          & 0.020          & 0.010          & 0.020          & 0.017          \\ \cline{3-13} 
\multicolumn{2}{c|}{\multirow{2}{*}{dwtDctSvd}}                                                                                   & AUC                     & \textbf{1.000}         & 0.949          & 0.286          & 0.600          & 0.649          & 0.621          & 0.797          & 0.597          & 0.487          & 0.627          \\
\multicolumn{2}{c|}{}                                                                                                             & T@1F                    & \textbf{1.000}         & 0.850          & 0.017          & 0.025          & 0.020          & 0.228          & 0.320          & 0.010          & 0.010          & 0.113          \\ \cline{3-13} 
\multicolumn{2}{c|}{\multirow{2}{*}{rivaGan}}                                                                                     & AUC                     & \textbf{1.000}         & \textbf{1.000} & 0.671          & 0.776          & 0.939          & 0.847          & 0.931          & 0.747          & 0.527          & 0.735          \\
\multicolumn{2}{c|}{}                                                                                                             & T@1F                    & \textbf{1.000}         & \textbf{1.000} & 0.657          & 0.600          & 0.780          & 0.759          & 0.510          & 0.150          & 0.050          & 0.237          \\ \cline{3-13} 
\multicolumn{2}{c|}{\multirow{2}{*}{watermarkDM}}                                                                                 & AUC                     & \textbf{1.000}         & \textbf{1.000} & 0.724          & 0.515          & 0.999          & 0.810          & 0.999          & 0.915          & 0.671          & 0.862          \\
\multicolumn{2}{c|}{}                                                                                                             & T@1F                    & \textbf{1.000}         & \textbf{1.000} & 0.656          & 0.112          & 0.991          & 0.690          & 0.920          & 0.340          & 0.000          & 0.420          \\ \cline{3-13} 
\multicolumn{2}{c|}{\multirow{2}{*}{SSL}}                                                                                         & AUC                     & \textbf{1.000}         & \textbf{1.000} & 0.992          & \textbf{0.991} & 0.621          & 0.901          & 0.959          & 0.695          & 0.750          & 0.801          \\
\multicolumn{2}{c|}{}                                                                                                             & T@1F                    & \textbf{1.000}         & \textbf{1.000} & 0.971          & \textbf{0.970} & 0.312          & 0.813          & 0.840          & 0.160          & 0.090          & 0.363          \\ \cline{3-13} 
\multicolumn{2}{c|}{\multirow{2}{*}{TreeRing}}                                                                                    & AUC                     & \textbf{1.000}         & 0.945          & 0.937          & 0.938          & 0.991          & 0.962          & \textbf{1.000} & 0.599        & 0.838          & 0.812          \\
\multicolumn{2}{c|}{}                                                                                                             & T@1F                    & \textbf{1.000}         & 0.902          & 0.747          & 0.708          & 0.952          & 0.862          & \textbf{1.000} & 0.000         & 0.150          & 0.383          \\ \hline
\multicolumn{1}{c|}{\multirow{6}{*}{\begin{tabular}[c]{@{}c@{}}VAR   \\      (Ours)\end{tabular}}}    & \multirow{2}{*}{LBW-Post} & AUC                     & 0.997                  & 0.988          & 0.989          & 0.659          & 0.981          & 0.923          & 0.997          & 0.933          & 0.650          & 0.860          \\
\multicolumn{1}{c|}{}                                                                                 &                           & T@1F                    & 0.980                  & 0.943          & 0.948          & 0.287          & 0.912          & 0.814          & 0.972          & 0.780          & 0.080          & 0.611          \\ \cline{3-13} 
\multicolumn{1}{c|}{}                                                                                 & \multirow{2}{*}{LBW-Hard}   & AUC                     & 0.999                  & 0.995          & 0.994          & 0.660          & 0.988          & 0.927          & 0.991          & 0.896          & 0.623          & 0.837          \\
\multicolumn{1}{c|}{}                                                                                 &                           & T@1F                    & 0.995                  & 0.967          & 0.969          & 0.281          & 0.932          & 0.829          & 0.903          & 0.526          & 0.000          & 0.476          \\ \cline{3-13} 
\multicolumn{1}{c|}{}                                                                                 & \multirow{2}{*}{LBX-Soft}      & AUC                     & \textbf{1.000}         & 0.995          & 0.994          & 0.665          & 0.989          & 0.929          & 0.995          & 0.892          & 0.626          & 0.838          \\
\multicolumn{1}{c|}{}                                                                                 &                           & T@1F                    & 0.997                  & 0.977          & 0.966          & 0.275          & 0.934          & 0.830          & 0.930          & 0.480          & 0.010          & 0.473          \\ \hline
\multicolumn{1}{c|}{\multirow{6}{*}{\begin{tabular}[c]{@{}c@{}}VQ-GAN   \\      (Ours)\end{tabular}}} & \multirow{2}{*}{LBW-Post} & AUC                     & 0.978                  & 0.972          & 0.856          & 0.773          & 0.948          & 0.905          & 0.969          & 0.922          & 0.665          & 0.852          \\
\multicolumn{1}{c|}{}                                                                                 &                           & T@1F                    & 0.760                  & 0.728          & 0.329          & 0.274          & 0.699          & 0.558          & 0.704          & 0.660          & 0.140          & 0.501          \\ \cline{3-13} 
\multicolumn{1}{c|}{}                                                                                 & \multirow{2}{*}{LBW-Hard}   & AUC                     & 0.998                  & 0.969          & 0.858          & 0.939          & 0.921          & 0.922          & 0.993          & 0.973          & 0.857          & 0.941          \\
\multicolumn{1}{c|}{}                                                                                 &                           & T@1F                    & 0.993                  & 0.909          & 0.700          & 0.641          & 0.745          & 0.749          & 0.978          & 0.870          & 0.370          & 0.739          \\ \cline{3-13} 
\multicolumn{1}{c|}{}                                                                                 & \multirow{2}{*}{LBX-Soft}      & AUC                     & 0.999                  & 0.990          & 0.934          & 0.966          & 0.995          & 0.977          & 0.998          & 0.994          & 0.915          & 0.969          \\
\multicolumn{1}{c|}{}                                                                                 &                           & T@1F                    & 0.998                  & 0.977          & 0.770          & 0.776          & 0.977          & 0.900          & 0.998          & 0.990          & 0.610          & 0.866          \\ \hline
\multicolumn{1}{c|}{\multirow{6}{*}{\begin{tabular}[c]{@{}c@{}}RAR   \\      (Ours)\end{tabular}}}    & \multirow{2}{*}{LBW-Post} & AUC                     & \textbf{1.000}         & \textbf{1.000} & 0.995          & \textbf{0.991} & 0.999          & \textbf{0.996} & \textbf{1.000} & 0.998          & \textbf{0.988} & \textbf{0.995} \\
\multicolumn{1}{c|}{}                                                                                 &                           & T@1F                    & \textbf{1.000}         & 0.999          & 0.956          & 0.918          & 0.995          & \textbf{0.967} & \textbf{1.000} & 0.960          & \textbf{0.850} & \textbf{0.937} \\ \cline{3-13} 
\multicolumn{1}{c|}{}                                                                                 & \multirow{2}{*}{LBW-Hard}   & AUC                     & \textbf{1.000}         & \textbf{1.000} & 0.997          & 0.964          & 0.999          & 0.990          & \textbf{1.000} & \textbf{1.000} & 0.978          & 0.993          \\
\multicolumn{1}{c|}{}                                                                                 &                           & T@1F                    & \textbf{1.000}         & 0.998          & 0.973          & 0.846          & 0.993          & 0.953          & \textbf{1.000} & \textbf{1.000} & 0.800          & 0.933          \\ \cline{3-13} 
\multicolumn{1}{c|}{}                                                                                 & \multirow{2}{*}{LBX-Soft}      & AUC                     & \textbf{1.000}         & \textbf{1.000} & \textbf{0.999} & 0.961          & \textbf{1.000} & 0.990          & \textbf{1.000} & \textbf{1.000} & 0.978          & 0.993          \\
\multicolumn{1}{c|}{}                                                                                 &                           & T@1F                    & \textbf{1.000}         & \textbf{1.000} & \textbf{0.981} & 0.844          & \textbf{0.999} & 0.956          & \textbf{1.000} & 0.990          & 0.760          & 0.917          \\ \hline
\end{tabular}
\label{tab:main}
\end{table*}
Table~\ref{tab:main} presents a comprehensive evaluation of watermarking methods under conventional and regeneration attacks, where our approach demonstrates strong robustness across all attacks. Notably, LBW achieves state-of-the-art performance against regeneration attacks; for instance, LBW-Post on RAR attains an AUC of 0.995 and TPR@1FPR of 0.937, significantly outperforming WatermarkDM. LBW-Soft further surpasses LBW-Hard in robustness. While our method is effective across different AR architectures, VAR shows slightly reduced robustness. This can be attributed to the multiscale quantization process of VAR, where high-frequency information is primarily captured by the largest-scale token map\cite{lee2022rqvae,tian2025var}. CtrlRegen preserves semantic structures while suppressing fine-grained details, which makes large-scale token maps more vulnerable to such attacks.

\subsection{Ablation Studies}
\label{sec:ablation}
\begin{figure*}[tbp]
    \centering
    \includegraphics[width=1.0\textwidth]{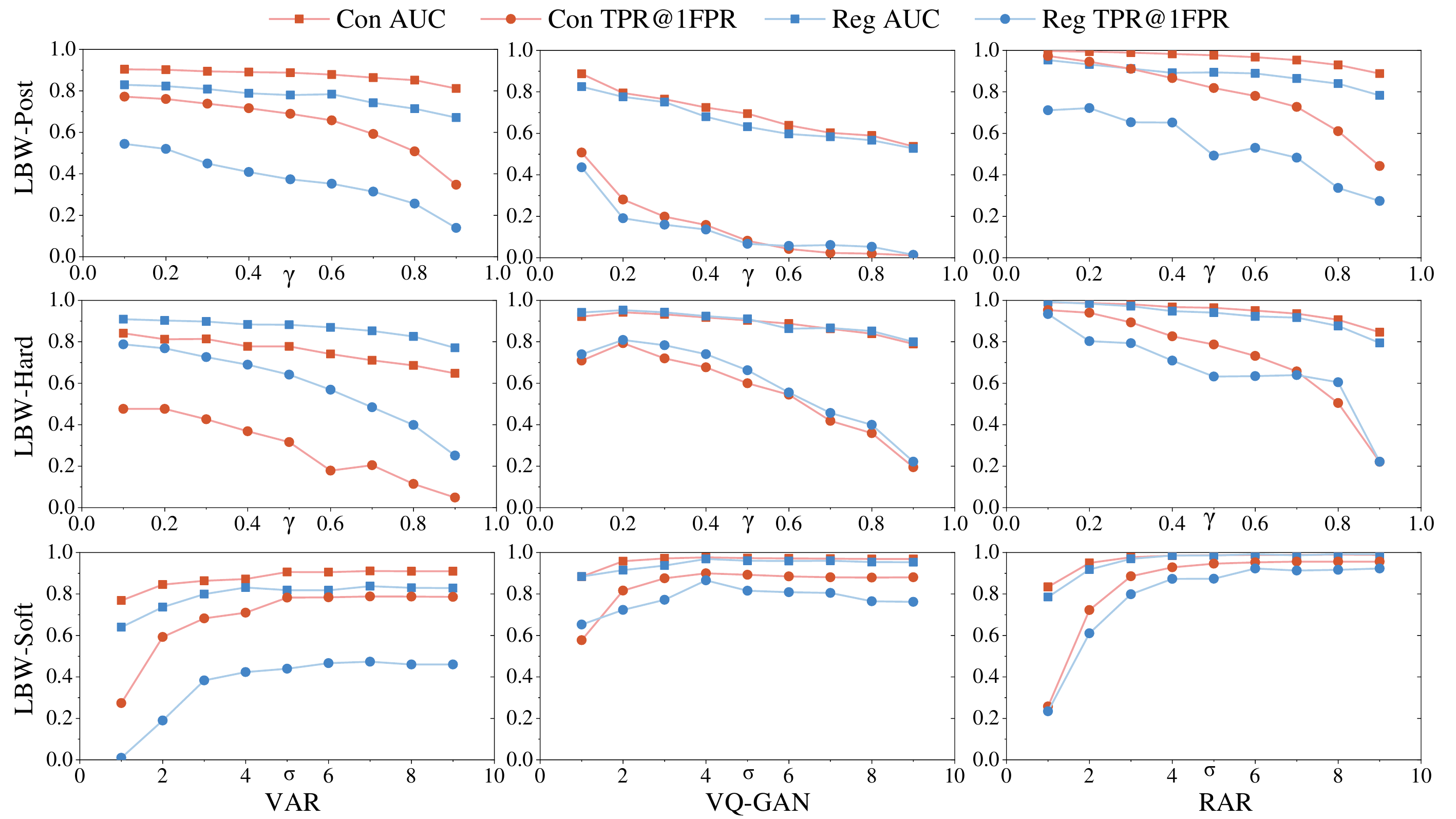}
    \caption{Impact of $\gamma$ and $\sigma$ in robustness against Conventional (Con) and Regeneration (Reg) attacks.}
    \label{fig:ablate_hyperparameter}
\end{figure*}
\begin{figure}[tbp]
    \centering
    \includegraphics[width=0.6\linewidth]{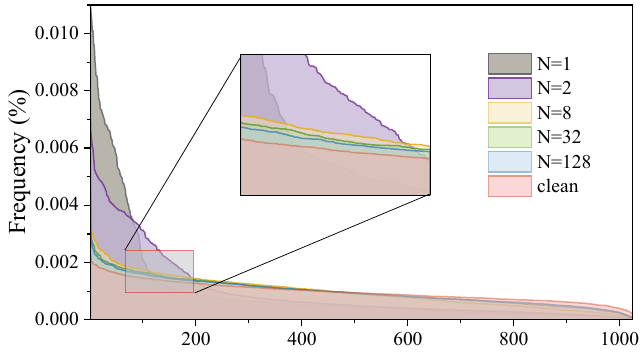}
    \caption{Comparison of token frequency distributions with varying green list number $N$.}
    \label{fig:token_frequency}
\end{figure}

Figures~\ref{fig:ablate_hyperparameter} (top rows) show that the robustness of LBW-Post and LBW-Hard generally improves as $\gamma$ decreases. An exception occurs for LBW-Hard on VQ-GAN, where $\gamma = 0.2$ outperforms $\gamma = 0.1$. This is because at very low $\gamma$ values, the model frequently fails to generate images, leading to reduced token consistency and sub-optimal robustness (see appendix~\ref{sec:token consistency} for detailed analyses). Based on these findings, we set the default values of $\gamma$ to 0.2, 0.1, and 0.1 for VQ-GAN, VAR, and RAR, respectively. Note that when evaluating the effect of $\sigma$ for LBW-Soft, the $\gamma$ is set to the default values. As shown in the bottom row of Figure~\ref{fig:ablate_hyperparameter}, the robustness of LBW-Soft initially improves with increasing $\sigma$ and eventually saturated. The results indicate that robustness initially improves with increasing values of $\sigma$ and eventually becomes saturated. While larger $\sigma$ values enhance watermark detectability, excessively high values can compromise image quality by overly restricting token generation. To achieve a balance between robustness and image fidelity, we adopt $\sigma=7,4$, and $8$ for VAR, VQ-GAN, and RAR. To prevent green tokens from being inferred, we introduce $N$ multiple green lists for watermark embedding. Using RAR with $\gamma=0.1$, we vary $N\in\{1,2,8,32,128\}$ and generate 10,000 post-hoc watermarked images per setting to study the impact of $N$ on the token frequency distribution of watermarked images. Figure~\ref{fig:token_frequency} shows that when $N$ is small (e.g., 1 or 2), the token frequency distribution deviates significantly from that of clean images, enabling white-box attacks. However, as $N$ increases to 32 and beyond, the token frequency distribution converges closely to that of clean images, effectively eliminating distinguishable statistical cues and rendering frequency-based attacks. Consequently, we choose $N=32$ in this work.

\textbf{More experiments, including visual quality analysis, token consistency analysis or other visual results can be found in the appendix.}

\section{Conclusion}
In this work, we present LBW, a novel watermarking framework for AR-based image generation that introduces a controlled bias in token selection during the generation process to embed a robust and detectable watermark. Our method can be seamlessly integrated into current AR image generation pipelines, achieving state-of-the-art robustness against regeneration attacks. Additionally, we extend LBW to a post-hoc watermarking scheme, showcasing its adaptability in both in-generation and post-hoc scenarios. The multiple green list strategy is further introduced to enhance robustness against white-box attacks.

\bibliographystyle{unsrt}  
\bibliography{main}  %%% Remove comment to use the external .bib file (using bibtex).

\begin{thebibliography}{10}

\bibitem{sdxl}
Dustin Podell, Zion English, Kyle Lacey, Andreas Blattmann, Tim Dockhorn, Jonas M{\"u}ller, Joe Penna, and Robin Rombach.
\newblock Sdxl: Improving latent diffusion models for high-resolution image synthesis.
\newblock {\em arXiv preprint arXiv:2307.01952}, 2023.

\bibitem{sd3}
Patrick Esser, Sumith Kulal, Andreas Blattmann, Rahim Entezari, Jonas M{\"u}ller, Harry Saini, Yam Levi, Dominik Lorenz, Axel Sauer, Frederic Boesel, et~al.
\newblock Scaling rectified flow transformers for high-resolution image synthesis.
\newblock In {\em Forty-first International Conference on Machine Learning}, 2024.

\bibitem{ldm}
Robin Rombach, Andreas Blattmann, Dominik Lorenz, Patrick Esser, and Bj{\"o}rn Ommer.
\newblock High-resolution image synthesis with latent diffusion models.
\newblock In {\em Proceedings of the IEEE/CVF conference on computer vision and pattern recognition}, pages 10684--10695, 2022.

\bibitem{zhang2024ssr}
Yuxuan Zhang, Yiren Song, Jiaming Liu, Rui Wang, Jinpeng Yu, Hao Tang, Huaxia Li, Xu~Tang, Yao Hu, Han Pan, et~al.
\newblock Ssr-encoder: Encoding selective subject representation for subject-driven generation.
\newblock In {\em Proceedings of the IEEE/CVF Conference on Computer Vision and Pattern Recognition}, pages 8069--8078, 2024.

\bibitem{zhang2024fast}
Yuxuan Zhang, Yiren Song, Jinpeng Yu, Han Pan, and Zhongliang Jing.
\newblock Fast personalized text to image synthesis with attention injection.
\newblock In {\em ICASSP 2024-2024 IEEE International Conference on Acoustics, Speech and Signal Processing (ICASSP)}, pages 6195--6199. IEEE, 2024.

\bibitem{zhang2025easycontrol}
Yuxuan Zhang, Yirui Yuan, Yiren Song, Haofan Wang, and Jiaming Liu.
\newblock Easycontrol: Adding efficient and flexible control for diffusion transformer.
\newblock {\em arXiv preprint arXiv:2503.07027}, 2025.

\bibitem{yang2024editworld}
Ling Yang, Bohan Zeng, Jiaming Liu, Hong Li, Minghao Xu, Wentao Zhang, and Shuicheng Yan.
\newblock Editworld: Simulating world dynamics for instruction-following image editing.
\newblock {\em arXiv preprint arXiv:2405.14785}, 2024.

\bibitem{huang2025photodoodle}
Shijie Huang, Yiren Song, Yuxuan Zhang, Hailong Guo, Xueyin Wang, Mike~Zheng Shou, and Jiaming Liu.
\newblock Photodoodle: Learning artistic image editing from few-shot pairwise data.
\newblock {\em arXiv preprint arXiv:2502.14397}, 2025.

\bibitem{shi2024seededit}
Yichun Shi, Peng Wang, and Weilin Huang.
\newblock Seededit: Align image re-generation to image editing.
\newblock {\em arXiv preprint arXiv:2411.06686}, 2024.

\bibitem{kawar2023imagic}
Bahjat Kawar, Shiran Zada, Oran Lang, Omer Tov, Huiwen Chang, Tali Dekel, Inbar Mosseri, and Michal Irani.
\newblock Imagic: Text-based real image editing with diffusion models.
\newblock In {\em Proceedings of the IEEE/CVF conference on computer vision and pattern recognition}, pages 6007--6017, 2023.

\bibitem{stablemakeup}
Yuxuan Zhang, Lifu Wei, Qing Zhang, Yiren Song, Jiaming Liu, Huaxia Li, Xu~Tang, Yao Hu, and Haibo Zhao.
\newblock Stable-makeup: When real-world makeup transfer meets diffusion model.
\newblock {\em arXiv preprint arXiv:2403.07764}, 2024.

\bibitem{zhang2025stable}
Yuxuan Zhang, Qing Zhang, Yiren Song, Jichao Zhang, Hao Tang, and Jiaming Liu.
\newblock Stable-hair: Real-world hair transfer via diffusion model.
\newblock In {\em Proceedings of the AAAI Conference on Artificial Intelligence}, volume~39, pages 10348--10356, 2025.

\bibitem{zhang2024show}
David~Junhao Zhang, Jay~Zhangjie Wu, Jia-Wei Liu, Rui Zhao, Lingmin Ran, Yuchao Gu, Difei Gao, and Mike~Zheng Shou.
\newblock Show-1: Marrying pixel and latent diffusion models for text-to-video generation.
\newblock {\em International Journal of Computer Vision}, pages 1--15, 2024.

\bibitem{song2025makeanything}
Yiren Song, Cheng Liu, and Mike~Zheng Shou.
\newblock Makeanything: Harnessing diffusion transformers for multi-domain procedural sequence generation.
\newblock {\em arXiv preprint arXiv:2502.01572}, 2025.

\bibitem{song2024processpainter}
Yiren Song, Shijie Huang, Chen Yao, Xiaojun Ye, Hai Ci, Jiaming Liu, Yuxuan Zhang, and Mike~Zheng Shou.
\newblock Processpainter: Learn painting process from sequence data.
\newblock {\em arXiv preprint arXiv:2406.06062}, 2024.

\bibitem{wan2024grid}
Cong Wan, Xiangyang Luo, Zijian Cai, Yiren Song, Yunlong Zhao, Yifan Bai, Yuhang He, and Yihong Gong.
\newblock Grid: Visual layout generation.
\newblock {\em arXiv preprint arXiv:2412.10718}, 2024.

\bibitem{flux2024}
Black~Forest Labs.
\newblock Flux.
\newblock \url{https://github.com/black-forest-labs/flux}, 2024.

\bibitem{betker2023improving}
James Betker, Gabriel Goh, Li~Jing, Tim Brooks, Jianfeng Wang, Linjie Li, Long Ouyang, Juntang Zhuang, Joyce Lee, Yufei Guo, et~al.
\newblock Improving image generation with better captions.
\newblock {\em Computer Science. https://cdn. openai. com/papers/dall-e-3. pdf}, 2(3):8, 2023.

\bibitem{tian2025var}
Keyu Tian, Yi~Jiang, Zehuan Yuan, Bingyue Peng, and Liwei Wang.
\newblock Visual autoregressive modeling: Scalable image generation via next-scale prediction.
\newblock {\em Advances in neural information processing systems}, 37:84839--84865, 2025.

\bibitem{han2024infinity}
Jian Han, Jinlai Liu, Yi~Jiang, Bin Yan, Yuqi Zhang, Zehuan Yuan, Bingyue Peng, and Xiaobing Liu.
\newblock Infinity: Scaling bitwise autoregressive modeling for high-resolution image synthesis.
\newblock {\em arXiv preprint arXiv:2412.04431}, 2024.

\bibitem{yu2024rar}
Qihang Yu, Ju~He, Xueqing Deng, Xiaohui Shen, and Liang-Chieh Chen.
\newblock Randomized autoregressive visual generation.
\newblock {\em arXiv preprint arXiv:2411.00776}, 2024.

\bibitem{xie2024showo}
Jinheng Xie, Weijia Mao, Zechen Bai, David~Junhao Zhang, Weihao Wang, Kevin~Qinghong Lin, Yuchao Gu, Zhijie Chen, Zhenheng Yang, and Mike~Zheng Shou.
\newblock Show-o: One single transformer to unify multimodal understanding and generation, 2024.

\bibitem{zhou2024transfusionpredicttokendiffuse}
Chunting Zhou, Lili Yu, Arun Babu, Kushal Tirumala, Michihiro Yasunaga, Leonid Shamis, Jacob Kahn, Xuezhe Ma, Luke Zettlemoyer, and Omer Levy.
\newblock Transfusion: Predict the next token and diffuse images with one multi-modal model, 2024.

\bibitem{Rahman_2013}
Md.~Maklachur Rahman.
\newblock A dwt, dct and svd based watermarking technique to protect the image piracy.
\newblock {\em International Journal of Managing Public Sector Information and Communication Technologies}, 4(2):21–32, June 2013.

\bibitem{zhang2019rivagan}
Kevin~Alex Zhang, Lei Xu, Alfredo Cuesta-Infante, and Kalyan Veeramachaneni.
\newblock Robust invisible video watermarking with attention, 2019.

\bibitem{fernandez2022watermarking}
Pierre Fernandez, Alexandre Sablayrolles, Teddy Furon, Herv{\'e} J{\'e}gou, and Matthijs Douze.
\newblock Watermarking images in self-supervised latent spaces.
\newblock In {\em ICASSP 2022-2022 IEEE International Conference on Acoustics, Speech and Signal Processing (ICASSP)}, pages 3054--3058. IEEE, 2022.

\bibitem{wen2023treering}
Yuxin Wen, John Kirchenbauer, Jonas Geiping, and Tom Goldstein.
\newblock Tree-ring watermarks: Fingerprints for diffusion images that are invisible and robust.
\newblock {\em arXiv preprint arXiv:2305.20030}, 2023.

\bibitem{ci2024ringid}
Hai Ci, Pei Yang, Yiren Song, and Mike~Zheng Shou.
\newblock Ringid: Rethinking tree-ring watermarking for enhanced multi-key identification.
\newblock In {\em European Conference on Computer Vision}, pages 338--354. Springer, 2024.

\bibitem{huang2024robinrobustinvisiblewatermarks}
Huayang Huang, Yu~Wu, and Qian Wang.
\newblock Robin: Robust and invisible watermarks for diffusion models with adversarial optimization, 2024.

\bibitem{nie2022diffusionmodelsadversarialpurification}
Weili Nie, Brandon Guo, Yujia Huang, Chaowei Xiao, Arash Vahdat, and Anima Anandkumar.
\newblock Diffusion models for adversarial purification, 2022.

\bibitem{liu2024ctrlregen}
Yepeng Liu, Yiren Song, Hai Ci, Yu~Zhang, Haofan Wang, Mike~Zheng Shou, and Yuheng Bu.
\newblock Image watermarks are removable using controllable regeneration from clean noise, 2024.

\bibitem{kirchenbauer2023reliability}
John Kirchenbauer, Jonas Geiping, Yuxin Wen, Manli Shu, Khalid Saifullah, Kezhi Kong, Kasun Fernando, Aniruddha Saha, Micah Goldblum, and Tom Goldstein.
\newblock On the reliability of watermarks for large language models.
\newblock {\em arXiv preprint arXiv:2306.04634}, 2023.

\bibitem{hu2023unbiased}
Zhengmian Hu, Lichang Chen, Xidong Wu, Yihan Wu, Hongyang Zhang, and Heng Huang.
\newblock Unbiased watermark for large language models.
\newblock {\em arXiv preprint arXiv:2310.10669}, 2023.

\bibitem{pang2024randardecoderonlyautoregressivevisual}
Ziqi Pang, Tianyuan Zhang, Fujun Luan, Yunze Man, Hao Tan, Kai Zhang, William~T. Freeman, and Yu-Xiong Wang.
\newblock Randar: Decoder-only autoregressive visual generation in random orders, 2024.

\bibitem{LSB}
R.B. Wolfgang and E.J. Delp.
\newblock A watermark for digital images.
\newblock In {\em Proceedings of 3rd IEEE International Conference on Image Processing}, volume~3, pages 219--222 vol.3, 1996.

\bibitem{dwtdctsvd}
K.~A. Navas, Mathews~Cheriyan Ajay, M.~Lekshmi, Tampy~S. Archana, and M.~Sasikumar.
\newblock Dwt-dct-svd based watermarking.
\newblock In {\em 2008 3rd International Conference on Communication Systems Software and Middleware and Workshops (COMSWARE '08)}, pages 271--274, 2008.

\bibitem{fernandez2022watermarkingimagesselfsupervisedlatent}
Pierre Fernandez, Alexandre Sablayrolles, Teddy Furon, Hervé Jégou, and Matthijs Douze.
\newblock Watermarking images in self-supervised latent spaces, 2022.

\bibitem{Tancik2020stegastamp}
Matthew Tancik, Ben Mildenhall, and Ren Ng.
\newblock Stegastamp: Invisible hyperlinks in physical photographs.
\newblock In {\em Proceedings of the IEEE/CVF Conference on Computer Vision and Pattern Recognition (CVPR)}, June 2020.

\bibitem{Fernandez_2023_stablesignature}
Pierre Fernandez, Guillaume Couairon, Herv\'e J\'egou, Matthijs Douze, and Teddy Furon.
\newblock The stable signature: Rooting watermarks in latent diffusion models.
\newblock In {\em Proceedings of the IEEE/CVF International Conference on Computer Vision (ICCV)}, pages 22466--22477, October 2023.

\bibitem{ci2024wmadapter}
Hai Ci, Yiren Song, Pei Yang, Jinheng Xie, and Mike~Zheng Shou.
\newblock Wmadapter: Adding watermark control to latent diffusion models, 2024.

\bibitem{kirchenbauer2023watermark}
John Kirchenbauer, Jonas Geiping, Yuxin Wen, Jonathan Katz, Ian Miers, and Tom Goldstein.
\newblock A watermark for large language models.
\newblock In {\em International Conference on Machine Learning}, pages 17061--17084. PMLR, 2023.

\bibitem{lu2024entropy}
Yijian Lu, Aiwei Liu, Dianzhi Yu, Jingjing Li, and Irwin King.
\newblock An entropy-based text watermarking detection method.
\newblock {\em arXiv preprint arXiv:2403.13485}, 2024.

\bibitem{liu2024adaptive}
Yepeng Liu and Yuheng Bu.
\newblock Adaptive text watermark for large language models.
\newblock {\em arXiv preprint arXiv:2401.13927}, 2024.

\bibitem{lee2024wrotecodewatermarkingcode}
Taehyun Lee, Seokhee Hong, Jaewoo Ahn, Ilgee Hong, Hwaran Lee, Sangdoo Yun, Jamin Shin, and Gunhee Kim.
\newblock Who wrote this code? watermarking for code generation, 2024.

\bibitem{wouters2023optimizing}
Bram Wouters.
\newblock Optimizing watermarks for large language models.
\newblock {\em arXiv preprint arXiv:2312.17295}, 2023.

\bibitem{liu2024semanticinvariantrobustwatermark}
Aiwei Liu, Leyi Pan, Xuming Hu, Shiao Meng, and Lijie Wen.
\newblock A semantic invariant robust watermark for large language models, 2024.

\bibitem{he2024can}
Zhiwei He, Binglin Zhou, Hongkun Hao, Aiwei Liu, Xing Wang, Zhaopeng Tu, Zhuosheng Zhang, and Rui Wang.
\newblock Can watermarks survive translation? on the cross-lingual consistency of text watermark for large language models.
\newblock {\em arXiv preprint arXiv:2402.14007}, 2024.

\bibitem{chen2020generative}
Mark Chen, Alec Radford, Rewon Child, Jeffrey Wu, Heewoo Jun, David Luan, and Ilya Sutskever.
\newblock Generative pretraining from pixels.
\newblock In {\em International conference on machine learning}, pages 1691--1703. PMLR, 2020.

\bibitem{gregor2014deep}
Karol Gregor, Ivo Danihelka, Andriy Mnih, Charles Blundell, and Daan Wierstra.
\newblock Deep autoregressive networks.
\newblock In {\em International Conference on Machine Learning}, pages 1242--1250. PMLR, 2014.

\bibitem{parmar2018image}
Niki Parmar, Ashish Vaswani, Jakob Uszkoreit, Lukasz Kaiser, Noam Shazeer, Alexander Ku, and Dustin Tran.
\newblock Image transformer.
\newblock In {\em International conference on machine learning}, pages 4055--4064. PMLR, 2018.

\bibitem{oord2018neuraldiscreterepresentationlearning}
Aaron van~den Oord, Oriol Vinyals, and Koray Kavukcuoglu.
\newblock Neural discrete representation learning, 2018.

\bibitem{esser2021taming}
Patrick Esser, Robin Rombach, and Bjorn Ommer.
\newblock Taming transformers for high-resolution image synthesis.
\newblock In {\em Proceedings of the IEEE/CVF conference on computer vision and pattern recognition}, pages 12873--12883, 2021.

\bibitem{razavi2019generating}
Ali Razavi, Aaron Van~den Oord, and Oriol Vinyals.
\newblock Generating diverse high-fidelity images with vq-vae-2.
\newblock {\em Advances in neural information processing systems}, 32, 2019.

\bibitem{lee2022rqvae}
Doyup Lee, Chiheon Kim, Saehoon Kim, Minsu Cho, and Wook-Shin Han.
\newblock Autoregressive image generation using residual quantization.
\newblock In {\em Proceedings of the IEEE/CVF Conference on Computer Vision and Pattern Recognition}, pages 11523--11532, 2022.

\bibitem{van2017vqvae}
Aaron Van Den~Oord, Oriol Vinyals, et~al.
\newblock Neural discrete representation learning.
\newblock {\em Advances in neural information processing systems}, 30, 2017.

\bibitem{al2007combined}
Ali Al-Haj.
\newblock Combined dwt-dct digital image watermarking.
\newblock {\em Journal of computer science}, 3(9):740--746, 2007.

\bibitem{zhao2023recipe}
Yunqing Zhao, Tianyu Pang, Chao Du, Xiao Yang, Ngai-Man Cheung, and Min Lin.
\newblock A recipe for watermarking diffusion models.
\newblock {\em arXiv preprint arXiv:2303.10137}, 2023.

\end{thebibliography}

%%%%%%%%%%%%%%%%%%%%%%%%%%%%%%%%%%%%%%%%%%%%%%%%%%%%%%%%%%%%
\appendix

\section{Visual Quality Analysis}
This section presents a comprehensive analysis of the visual quality of the proposed watermarking methods—LBW-Post, LBW-Hard, and LBW-Soft—through both quantitative and qualitative evaluations.

\begin{figure}[tbp]
    \centering
    \includegraphics[width=\linewidth]{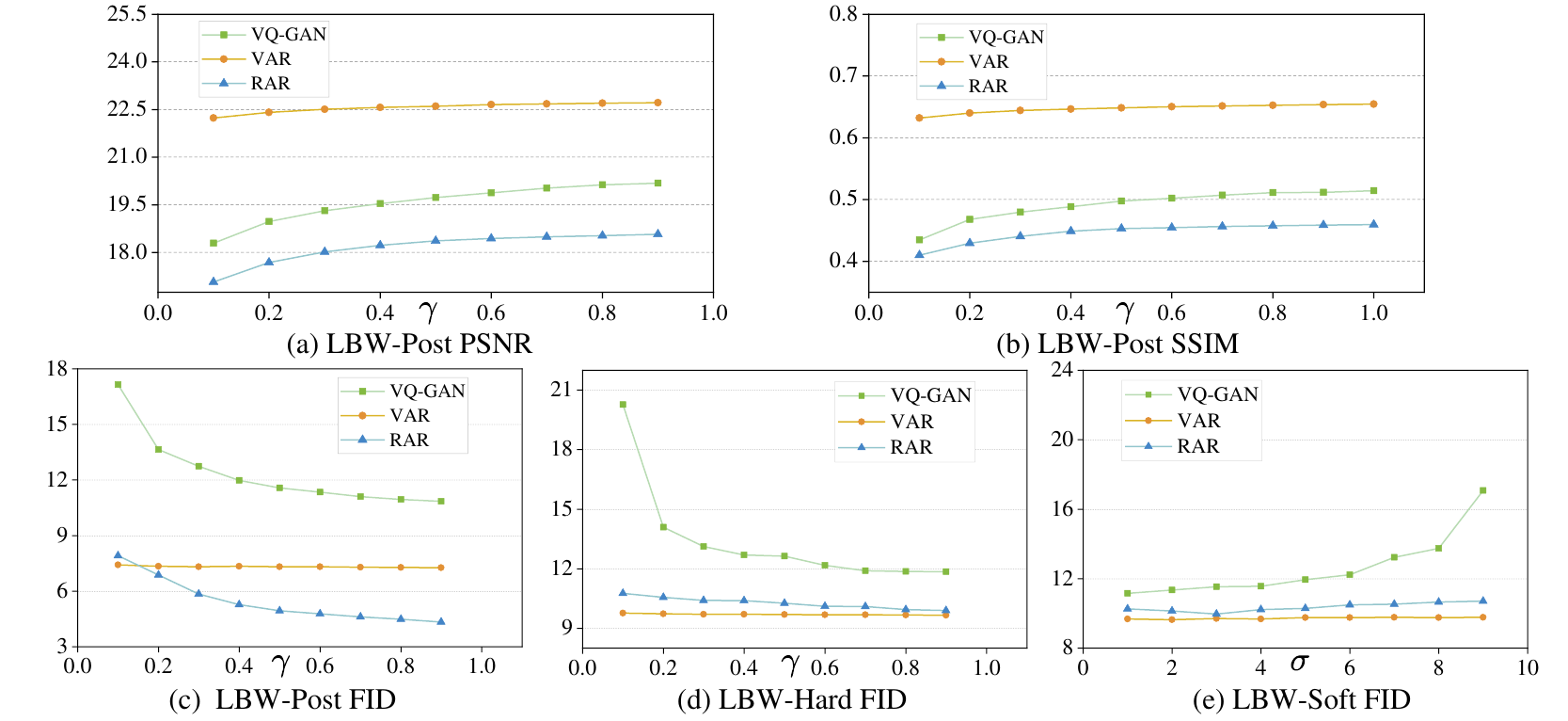}
    \caption{\textbf{Quantitative evaluation of visual quality for our LBW with varying $\gamma$ and $\sigma$}. Subfigures (a), (b), and (c) present PSNR, SSIM, and FID metrics for the LBW-Post watermark applied on VQ-GAN, VAR, and RAR models, while (d) and (e) show FID for our LBW-Hard and LBW-Soft, respectively.}
    \label{fig:visual_quantity}
\end{figure}

Figure~\ref{fig:visual_quantity} reports FID scores for our watermark methods with varying $\gamma$ and $\sigma$. PSNR and SSIM are reported exclusively for LBW-Post, as it is the only method with access to ground-truth images. The results indicate that increasing $\gamma$ consistently enhances image quality across all models, evidenced by improved PSNR and SSIM scores for LBW-Post, and by reduced FID values of both LBW-Post and LBW-Hard. Conversely, for LBW-Soft, FID scores increase with larger $\sigma$, indicating a degradation in perceptual quality. Compared to LBW-Hard at $\gamma=0.1$, LBW-Soft achieves lower FID scores across a large range of $\sigma$, reflecting its superior capability to balance watermark robustness with image fidelity. Notably, compared to VQ-GAN and RAR, VAR exhibits superior robustness to variations in both $\gamma$ and $\sigma$. This robustness stems from watermark embedding exclusively within its largest-scale token map, which encodes rich high-frequency information, thereby mitigating perceptual degradation. 

\textbf{Quantitative results}

\textbf{Qualitative results}
\begin{figure}[tbp]
    \centering
    \includegraphics[width=\linewidth]{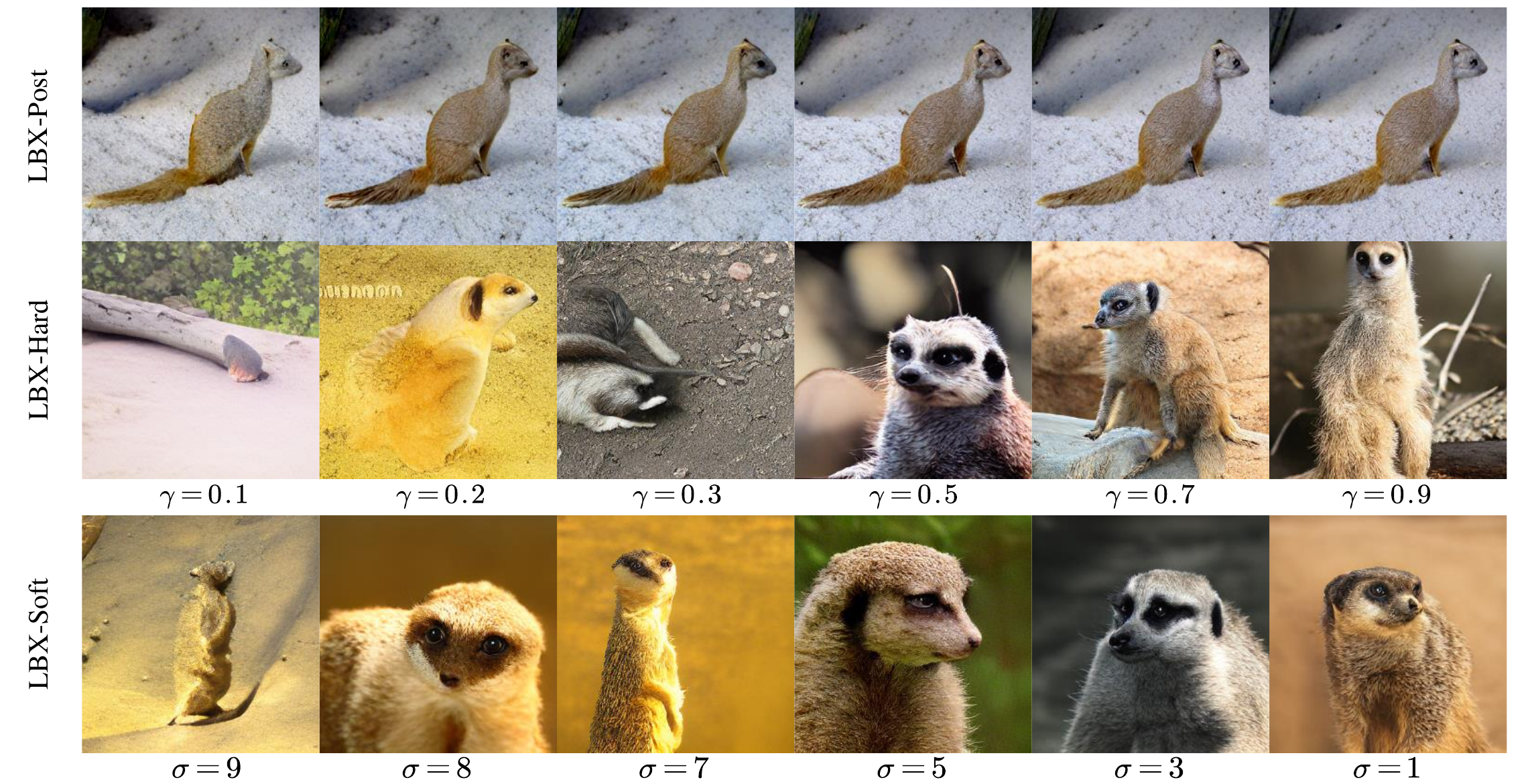}
    \caption{\textbf{Qualitative evaluation of visual quality for our LBW with varying $\gamma$ and $\sigma$}}
    \label{fig:visual_quality}
\end{figure}
Figure~\ref{fig:visual_quality} offers a visual comparison of watermarked images generated by VQ-GAN under the three watermarking schemes. The first two rows illustrate the results of LBW-Post and LBW-Hard, respectively, while the last presents the results of LBW-Soft. The image quality improves with increasing $\gamma$ for both LBW-Post and LBW-Hard. For LBW-Soft, reducing the noise parameter $\sigma$ enhances image quality. When $\gamma$ is low (e.g., 0.1), the LBW-Hard often suffers from degraded image quality and occasionally fails to produce class-relevant images. This limitation arises because LBW-Hard enforces strict token selection constrained solely to the green list, which significantly restricts the model’s expressive capacity. In contrast, LBW-Soft permits sampling outside the green list with moderated bias, thereby demonstrating greater robustness and superior visual quality under low green token ratios.

In summary, these quantitative and qualitative analyses corroborate that higher green token ratios $\gamma$ and lower logit bias constant $\sigma$ correlate with improved image quality, and our LBW-Soft effectively achieves a trade-off between watermark robustness and visual fidelity.

\section{Token Consistency}
\label{sec:token consistency}
\begin{figure}[tbp]
    \centering
    \includegraphics[width=0.7\linewidth]{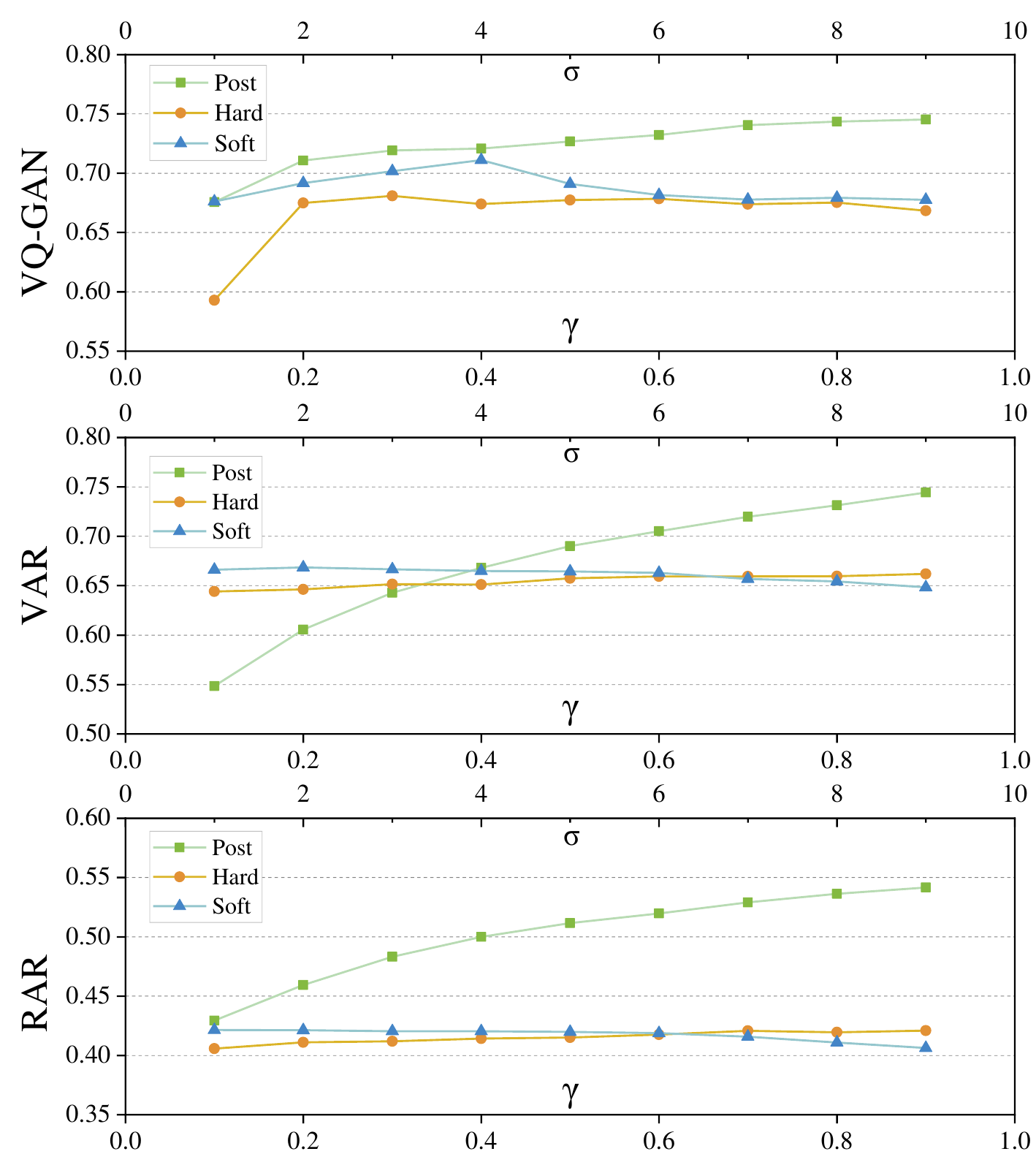}
    \caption{Token consistency with varying $\gamma$.}
    \label{fig:token_consistency}
\end{figure}
Figure~\ref{fig:token_consistency} presents a comparative analysis of the token consistency of our method for VQ-GAN, VAR, and RAR under different green list ratios $\gamma$ and logits biasing constant $\sigma$. Overall, the token consistency of the LBW-Post method increases with $\gamma$, whereas in-generation methods (LBW-Hard \& LBW-Soft) are insensitive to $\gamma$ and $\sigma$. This suggests that in-generation watermarking could improve detectability by increasing $\gamma$ and $\sigma$ without worrying about the token exchanges during detection.

\section{Visual Comparison between LBW-Hard and LBW-Soft}
\begin{figure}[tbp]
    \centering
    \includegraphics[width=\linewidth]{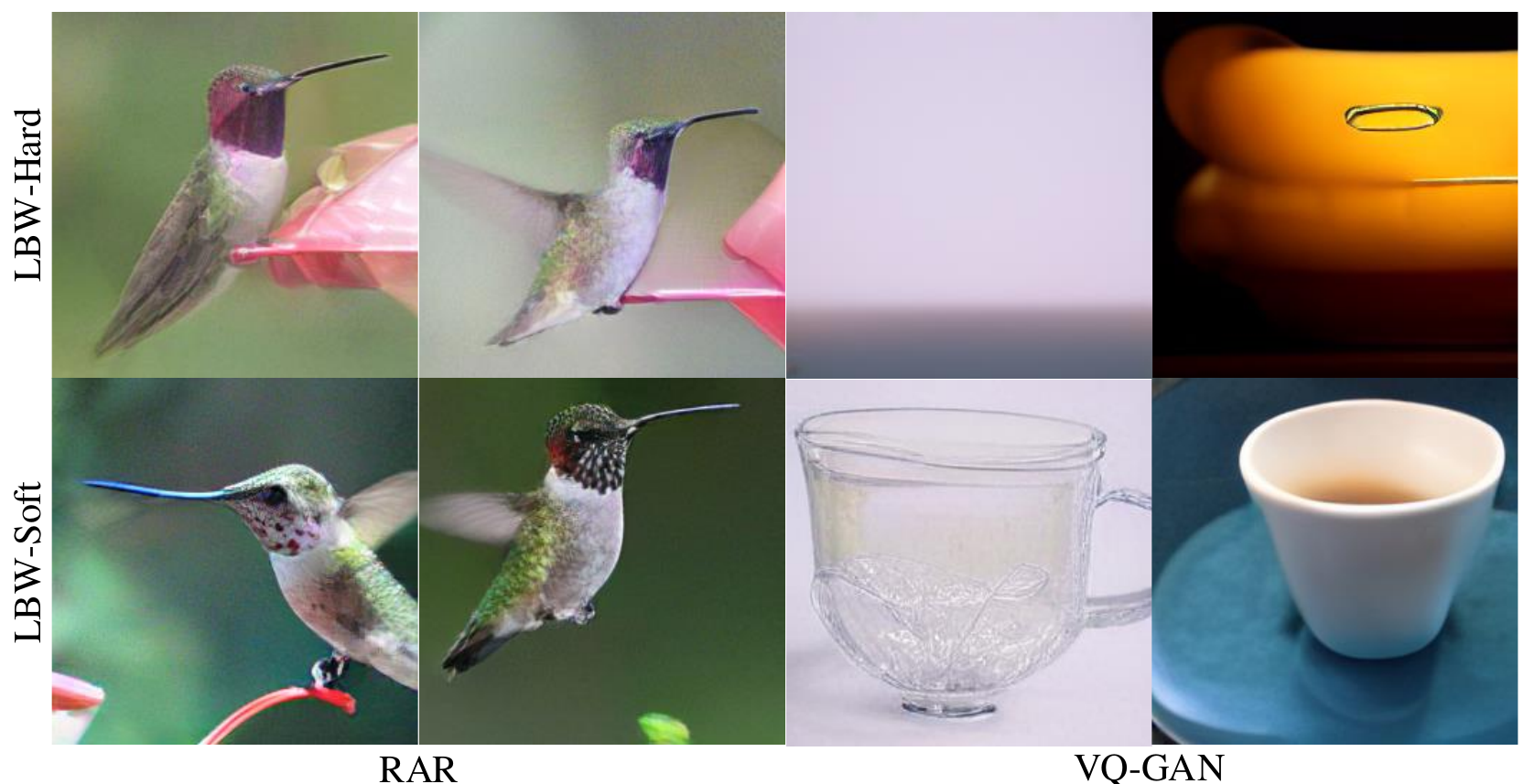}
    \caption{Comparison between watermark images produced by LBW-Hard and LBW-Soft.}
    \label{fig:simple_vs_lbw}
\end{figure}

Figure~\ref{fig:simple_vs_lbw} presents a comparative analysis of watermarked images synthesized using our LBW-Hard and LBW-Soft approaches, integrated with RAR and VQ-GAN. The results demonstrate that LBW-Hard results in reduced image quality when applied to RAR and even fails to generate class-relevant content in the case of VQ-GAN. In contrast, LBW-Soft produces images with finer details and realistic content. These findings highlight the efficacy of LBW-Soft in striking a balance between watermark robustness and image fidelity, making it a more suitable approach for AR image models.

\section{Numerical Robustness Results}
\label{sec:attacks}
To assess watermark robustness, we evaluate conventional and regeneration attacks. Conventional attacks include (1) \textbf{Gauss}ian attacks randomly apply Gaussian Noise with variation 0.1 and Gaussian Blur using an $8 \times 8$ filter) (2) \textbf{Color}Jitter perturbations involving randomly applying hue adjustments ($0.3$), saturation scaling ($3.0$), and contrast scaling ($3.0$), 
(3) \textbf{Geom}etric transformations (Crop\&Resize: 0.7, Random Rotation: $0^\circ$-$180^\circ$), and 
(4) \textbf{JPEG} compression (25\% ). Regeneration attacks include 
(1) \textbf{VAE} reconstruction via the VAE of Stable Diffusion 1.5, 
(2) \textbf{Diff}Pure~\cite{zhou2024transfusionpredicttokendiffuse} with timestep $t = 0.15$, and 
(3) \textbf{Ctrl}Regen~\cite{liu2024ctrlregen}.

In this section, we illustrate the numerical robustness of our proposed Lexical Bias Watermarking (LBW) methods—LBW-Post, LBW-Hard, and LBW-Soft—against the above watermark removal attacks in Table~\ref{post_hoc}, Table~\ref{hard}, and Table~\ref{soft}, respectively. The evaluation is conducted on three different AR image models: VAR, VQ-GAN, and RAR, across varying watermark embedding strengths, denoted by the parameter $\gamma$, ranging from 0.1 to 0.9. For LBW-Soft, we evaluate the effect of bias constant $\sigma$ under $\gamma=0.1,0.2, \ and \ 0.1$ for VAR, VQ-GAN, and RAR, respectively.  We assess performance using AUC (Area Under Curve) and T@1F (True Positive Rate at 1\% False Positive Rate), which indicate the detectability of the watermark under different attack conditions.

\begin{table*}[tbp]\small
\centering
\begin{tabular}{c|c|c|llllllll}
\hline
                         & $\gamma$               & Metric     & \multicolumn{1}{c}{Clean} & \multicolumn{1}{c}{Gaus} & \multicolumn{1}{c}{Color} & \multicolumn{1}{c}{Geom} & \multicolumn{1}{c}{JPEG} & \multicolumn{1}{c}{VAE} & \multicolumn{1}{c}{Diff} & \multicolumn{1}{c}{Ctrl} \\ \hline
\multirow{18}{*}{VAR}    & \multirow{2}{*}{0.100} & AUC        & 0.997                     & 0.988                    & 0.989                     & 0.659                    & 0.981                    & 0.997                   & 0.933                    & 0.650                    \\
                         &                        & { T@1F} & 0.980                     & 0.943                    & 0.948                     & 0.287                    & 0.912                    & 0.972                   & 0.780                    & 0.080                    \\
                         & \multirow{2}{*}{0.200} & AUC        & 0.998                     & 0.989                    & 0.985                     & 0.655                    & 0.977                    & 0.995                   & 0.923                    & 0.645                    \\
                         &                        & { T@1F} & 0.985                     & 0.940                    & 0.935                     & 0.280                    & 0.888                    & 0.968                   & 0.700                    & 0.100                    \\
                         & \multirow{2}{*}{0.300} & AUC        & 0.995                     & 0.983                    & 0.983                     & 0.639                    & 0.972                    & 0.996                   & 0.919                    & 0.614                    \\
                         &                        & { T@1F} & 0.973                     & 0.920                    & 0.906                     & 0.274                    & 0.852                    & 0.948                   & 0.580                    & 0.030                    \\
                         & \multirow{2}{*}{0.400} & AUC        & 0.994                     & 0.977                    & 0.977                     & 0.641                    & 0.967                    & 0.986                   & 0.868                    & 0.605                    \\
                         &                        & { T@1F} & 0.962                     & 0.876                    & 0.884                     & 0.275                    & 0.829                    & 0.932                   & 0.510                    & 0.000                    \\
                         & \multirow{2}{*}{0.500} & AUC        & 0.995                     & 0.974                    & 0.974                     & 0.645                    & 0.957                    & 0.984                   & 0.865                    & 0.588                    \\
                         &                        & { T@1F} & 0.961                     & 0.848                    & 0.867                     & 0.273                    & 0.772                    & 0.890                   & 0.410                    & 0.020                    \\
                         & \multirow{2}{*}{0.600} & AUC        & 0.991                     & 0.969                    & 0.967                     & 0.629                    & 0.950                    & 0.981                   & 0.886                    & 0.586                    \\
                         &                        & { T@1F} & 0.946                     & 0.802                    & 0.832                     & 0.267                    & 0.729                    & 0.892                   & 0.390                    & 0.010                    \\
                         & \multirow{2}{*}{0.700} & AUC        & 0.985                     & 0.951                    & 0.952                     & 0.627                    & 0.926                    & 0.975                   & 0.823                    & 0.540                    \\
                         &                        & { T@1F} & 0.911                     & 0.721                    & 0.745                     & 0.255                    & 0.646                    & 0.796                   & 0.350                    & 0.000                    \\
                         & \multirow{2}{*}{0.800} & AUC        & 0.979                     & 0.932                    & 0.934                     & 0.636                    & 0.906                    & 0.945                   & 0.765                    & 0.525                    \\
                         &                        & { T@1F} & 0.847                     & 0.606                    & 0.631                     & 0.249                    & 0.546                    & 0.638                   & 0.240                    & 0.020                    \\
                         & \multirow{2}{*}{0.900} & AUC        & 0.956                     & 0.883                    & 0.889                     & 0.617                    & 0.856                    & 0.881                   & 0.687                    & 0.516                    \\
                         &                        & { T@1F} & 0.659                     & 0.367                    & 0.451                     & 0.193                    & 0.378                    & 0.230                   & 0.050                    & 0.020                    \\ \hline
\multirow{18}{*}{VQ-GAN} & \multirow{2}{*}{0.100} & AUC        & 0.978                     & 0.972                    & 0.856                     & 0.773                    & 0.948                    & 0.969                   & 0.922                    & 0.665                    \\
                         &                        & { T@1F} & 0.760                     & 0.728                    & 0.329                     & 0.274                    & 0.699                    & 0.704                   & 0.660                    & 0.140                    \\
                         & \multirow{2}{*}{0.200} & AUC        & 0.898                     & 0.894                    & 0.749                     & 0.681                    & 0.856                    & 0.880                   & 0.850                    & 0.684                    \\
                         &                        & { T@1F} & 0.493                     & 0.433                    & 0.155                     & 0.171                    & 0.366                    & 0.434                   & 0.280                    & 0.010                    \\
                         & \multirow{2}{*}{0.300} & AUC        & 0.871                     & 0.864                    & 0.715                     & 0.661                    & 0.821                    & 0.866                   & 0.800                    & 0.687                    \\
                         &                        & { T@1F} & 0.369                     & 0.321                    & 0.112                     & 0.135                    & 0.226                    & 0.386                   & 0.240                    & 0.040                    \\
                         & \multirow{2}{*}{0.400} & AUC        & 0.829                     & 0.813                    & 0.676                     & 0.627                    & 0.784                    & 0.815                   & 0.749                    & 0.567                    \\
                         &                        & { T@1F} & 0.320                     & 0.264                    & 0.095                     & 0.102                    & 0.170                    & 0.262                   & 0.210                    & 0.040                    \\
                         & \multirow{2}{*}{0.500} & AUC        & 0.786                     & 0.774                    & 0.648                     & 0.617                    & 0.741                    & 0.771                   & 0.690                    & 0.510                    \\
                         &                        & { T@1F} & 0.169                     & 0.120                    & 0.053                     & 0.059                    & 0.097                    & 0.146                   & 0.100                    & 0.020                    \\
                         & \multirow{2}{*}{0.600} & AUC        & 0.721                     & 0.693                    & 0.606                     & 0.578                    & 0.678                    & 0.651                   & 0.631                    & 0.522                    \\
                         &                        & { T@1F} & 0.101                     & 0.068                    & 0.020                     & 0.023                    & 0.059                    & 0.056                   & 0.110                    & 0.020                    \\
                         & \multirow{2}{*}{0.700} & AUC        & 0.648                     & 0.648                    & 0.582                     & 0.556                    & 0.623                    & 0.605                   & 0.642                    & 0.508                    \\
                         &                        & { T@1F} & 0.049                     & 0.037                    & 0.015                     & 0.026                    & 0.015                    & 0.064                   & 0.110                    & 0.050                    \\
                         & \multirow{2}{*}{0.800} & AUC        & 0.629                     & 0.627                    & 0.566                     & 0.553                    & 0.610                    & 0.599                   & 0.596                    & 0.514                    \\
                         &                        & { T@1F} & 0.047                     & 0.031                    & 0.012                     & 0.017                    & 0.019                    & 0.040                   & 0.070                    & 0.070                    \\
                         & \multirow{2}{*}{0.900} & AUC        & 0.547                     & 0.554                    & 0.535                     & 0.514                    & 0.546                    & 0.510                   & 0.561                    & 0.483                    \\
                         &                        & { T@1F} & 0.016                     & 0.012                    & 0.009                     & 0.010                    & 0.015                    & 0.016                   & 0.000                    & 0.030                    \\ \hline
\multirow{18}{*}{RAR}    & \multirow{2}{*}{0.100} & AUC        & 1.000                     & 1.000                    & 0.995                     & 0.991                    & 0.999                    & 1.000                   & 0.993                    & 0.870                    \\
                         &                        & { T@1F} & 1.000                     & 0.999                    & 0.956                     & 0.918                    & 0.995                    & 1.000                   & 0.920                    & 0.240                    \\
                         & \multirow{2}{*}{0.200} & AUC        & 1.000                     & 1.000                    & 0.987                     & 0.984                    & 0.999                    & 1.000                   & 0.988                    & 0.815                    \\
                         &                        & { T@1F} & 1.000                     & 0.999                    & 0.897                     & 0.853                    & 0.980                    & 1.000                   & 0.940                    & 0.280                    \\
                         & \multirow{2}{*}{0.300} & AUC        & 1.000                     & 1.000                    & 0.977                     & 0.973                    & 0.998                    & 1.000                   & 0.981                    & 0.767                    \\
                         &                        & { T@1F} & 1.000                     & 0.998                    & 0.834                     & 0.764                    & 0.960                    & 1.000                   & 0.830                    & 0.220                    \\
                         & \multirow{2}{*}{0.400} & AUC        & 0.999                     & 0.999                    & 0.966                     & 0.959                    & 0.994                    & 1.000                   & 0.977                    & 0.715                    \\
                         &                        & { T@1F} & 0.999                     & 0.994                    & 0.753                     & 0.640                    & 0.947                    & 1.000                   & 0.940                    & 0.150                    \\
                         & \multirow{2}{*}{0.500} & AUC        & 0.999                     & 0.998                    & 0.959                     & 0.940                    & 0.988                    & 1.000                   & 0.975                    & 0.731                    \\
                         &                        & { T@1F} & 0.999                     & 0.986                    & 0.652                     & 0.553                    & 0.903                    & 1.000                   & 0.550                    & 0.110                    \\
                         & \multirow{2}{*}{0.600} & AUC        & 0.999                     & 0.997                    & 0.940                     & 0.918                    & 0.984                    & 1.000                   & 0.966                    & 0.734                    \\
                         &                        & { T@1F} & 0.999                     & 0.973                    & 0.583                     & 0.498                    & 0.849                    & 0.998                   & 0.710                    & 0.100                    \\
                         & \multirow{2}{*}{0.700} & AUC        & 1.000                     & 0.994                    & 0.912                     & 0.882                    & 0.980                    & 1.000                   & 0.947                    & 0.693                    \\
                         &                        & { T@1F} & 0.998                     & 0.951                    & 0.488                     & 0.384                    & 0.817                    & 0.990                   & 0.650                    & 0.070                    \\
                         & \multirow{2}{*}{0.800} & AUC        & 0.999                     & 0.983                    & 0.876                     & 0.834                    & 0.960                    & 0.998                   & 0.928                    & 0.660                    \\
                         &                        & { T@1F} & 0.987                     & 0.809                    & 0.370                     & 0.314                    & 0.571                    & 0.944                   & 0.380                    & 0.020                    \\
                         & \multirow{2}{*}{0.900} & AUC        & 0.996                     & 0.945                    & 0.817                     & 0.781                    & 0.902                    & 0.988                   & 0.878                    & 0.584                    \\
                         &                        & { T@1F} & 0.911                     & 0.485                    & 0.193                     & 0.281                    & 0.343                    & 0.812                   & 0.320                    & 0.060                    \\ \hline
\end{tabular}
\caption{Robustness for LBW-Post for VAR, VQ-GAN and RAR across different $\gamma$, ranging from 0.1 to 0.9.}
\label{post_hoc}
\end{table*}

\begin{table*}[tbp]\small
\centering
\begin{tabular}{c|c|c|llllllll}
\hline
                         & $\gamma$               & Metric     & \multicolumn{1}{c}{Clean} & \multicolumn{1}{c}{Gaus} & \multicolumn{1}{c}{Color} & \multicolumn{1}{c}{Geom} & \multicolumn{1}{c}{JPEG} & \multicolumn{1}{c}{VAE} & \multicolumn{1}{c}{Diff} & \multicolumn{1}{c}{Ctrl} \\ \hline
\multirow{18}{*}{VAR}    & \multirow{2}{*}{0.100} & AUC        & 0.999                     & 0.995                    & 0.994                     & 0.660                    & 0.988                    & 0.991                   & 0.896                    & 0.623                    \\
                         &                        & { T@1F} & 0.995                     & 0.967                    & 0.969                     & 0.281                    & 0.932                    & 0.903                   & 0.526                    & 0.000                    \\
                         & \multirow{2}{*}{0.200} & AUC        & 0.998                     & 0.992                    & 0.992                     & 0.645                    & 0.982                    & 0.983                   & 0.819                    & 0.542                    \\
                         &                        & { T@1F} & 0.988                     & 0.930                    & 0.958                     & 0.277                    & 0.909                    & 0.890                   & 0.500                    & 0.040                    \\
                         & \multirow{2}{*}{0.300} & AUC        & 0.997                     & 0.983                    & 0.985                     & 0.650                    & 0.972                    & 0.974                   & 0.844                    & 0.635                    \\
                         &                        & { T@1F} & 0.978                     & 0.903                    & 0.908                     & 0.269                    & 0.826                    & 0.838                   & 0.420                    & 0.020                    \\
                         & \multirow{2}{*}{0.400} & AUC        & 0.994                     & 0.975                    & 0.977                     & 0.629                    & 0.955                    & 0.954                   & 0.775                    & 0.638                    \\
                         &                        & { T@1F} & 0.958                     & 0.839                    & 0.888                     & 0.267                    & 0.764                    & 0.656                   & 0.330                    & 0.120                    \\
                         & \multirow{2}{*}{0.500} & AUC        & 0.992                     & 0.966                    & 0.973                     & 0.639                    & 0.951                    & 0.933                   & 0.792                    & 0.604                    \\
                         &                        & { T@1F} & 0.947                     & 0.775                    & 0.824                     & 0.260                    & 0.708                    & 0.519                   & 0.340                    & 0.090                    \\
                         & \multirow{2}{*}{0.600} & AUC        & 0.988                     & 0.953                    & 0.960                     & 0.637                    & 0.929                    & 0.910                   & 0.742                    & 0.572                    \\
                         &                        & { T@1F} & 0.912                     & 0.697                    & 0.769                     & 0.246                    & 0.564                    & 0.415                   & 0.100                    & 0.020                    \\
                         & \multirow{2}{*}{0.700} & AUC        & 0.978                     & 0.931                    & 0.941                     & 0.626                    & 0.912                    & 0.867                   & 0.725                    & 0.512                    \\
                         &                        & { T@1F} & 0.830                     & 0.557                    & 0.654                     & 0.229                    & 0.497                    & 0.273                   & 0.270                    & 0.070                    \\
                         & \multirow{2}{*}{0.800} & AUC        & 0.966                     & 0.898                    & 0.917                     & 0.619                    & 0.867                    & 0.828                   & 0.696                    & 0.609                    \\
                         &                        & { T@1F} & 0.752                     & 0.430                    & 0.531                     & 0.197                    & 0.437                    & 0.164                   & 0.110                    & 0.070                    \\
                         & \multirow{2}{*}{0.900} & AUC        & 0.925                     & 0.828                    & 0.853                     & 0.596                    & 0.808                    & 0.749                   & 0.684                    & 0.533                    \\
                         &                        & { T@1F} & 0.535                     & 0.243                    & 0.351                     & 0.144                    & 0.267                    & 0.076                   & 0.070                    & 0.000                    \\ \hline
\multirow{18}{*}{VQ-GAN} & \multirow{2}{*}{0.100} & AUC        & 0.998                     & 0.969                    & 0.858                     & 0.903                    & 0.921                    & 0.993                   & 0.973                    & 0.857                    \\
                         &                        & { T@1F} & 0.993                     & 0.909                    & 0.700                     & 0.482                    & 0.745                    & 0.978                   & 0.870                    & 0.370                    \\
                         & \multirow{2}{*}{0.200} & AUC        & 0.999                     & 0.987                    & 0.912                     & 0.939                    & 0.966                    & 0.998                   & 0.992                    & 0.869                    \\
                         &                        & { T@1F} & 0.998                     & 0.970                    & 0.714                     & 0.641                    & 0.849                    & 0.996                   & 0.990                    & 0.440                    \\
                         & \multirow{2}{*}{0.300} & AUC        & 1.000                     & 0.994                    & 0.902                     & 0.871                    & 0.964                    & 1.000                   & 0.984                    & 0.842                    \\
                         &                        & { T@1F} & 1.000                     & 0.979                    & 0.654                     & 0.411                    & 0.835                    & 0.999                   & 0.940                    & 0.410                    \\
                         & \multirow{2}{*}{0.400} & AUC        & 1.000                     & 0.990                    & 0.892                     & 0.831                    & 0.956                    & 1.000                   & 0.975                    & 0.796                    \\
                         &                        & { T@1F} & 0.999                     & 0.963                    & 0.547                     & 0.371                    & 0.827                    & 1.000                   & 0.890                    & 0.330                    \\
                         & \multirow{2}{*}{0.500} & AUC        & 1.000                     & 0.991                    & 0.881                     & 0.793                    & 0.950                    & 0.999                   & 0.967                    & 0.764                    \\
                         &                        & { T@1F} & 1.000                     & 0.953                    & 0.449                     & 0.339                    & 0.659                    & 0.997                   & 0.820                    & 0.170                    \\
                         & \multirow{2}{*}{0.600} & AUC        & 1.000                     & 0.986                    & 0.854                     & 0.779                    & 0.932                    & 0.998                   & 0.947                    & 0.644                    \\
                         &                        & { T@1F} & 0.997                     & 0.921                    & 0.367                     & 0.299                    & 0.591                    & 0.987                   & 0.660                    & 0.020                    \\
                         & \multirow{2}{*}{0.700} & AUC        & 0.998                     & 0.970                    & 0.820                     & 0.751                    & 0.908                    & 0.993                   & 0.913                    & 0.695                    \\
                         &                        & { T@1F} & 0.985                     & 0.794                    & 0.167                     & 0.292                    & 0.421                    & 0.859                   & 0.480                    & 0.030                    \\
                         & \multirow{2}{*}{0.800} & AUC        & 0.997                     & 0.959                    & 0.785                     & 0.736                    & 0.876                    & 0.987                   & 0.870                    & 0.698                    \\
                         &                        & { T@1F} & 0.974                     & 0.666                    & 0.168                     & 0.277                    & 0.327                    & 0.867                   & 0.251                    & 0.080                    \\
                         & \multirow{2}{*}{0.900} & AUC        & 0.989                     & 0.917                    & 0.727                     & 0.706                    & 0.809                    & 0.966                   & 0.792                    & 0.641                    \\
                         &                        & { T@1F} & 0.784                     & 0.376                    & 0.082                     & 0.227                    & 0.094                    & 0.544                   & 0.110                    & 0.010                    \\ \hline
\multirow{18}{*}{RAR}    & \multirow{2}{*}{0.100} & AUC        & 1.000                     & 1.000                    & 0.997                     & 0.964                    & 0.999                    & 1.000                   & 1.000                    & 0.978                    \\
                         &                        & { T@1F} & 1.000                     & 0.998                    & 0.973                     & 0.846                    & 0.993                    & 1.000                   & 1.000                    & 0.800                    \\
                         & \multirow{2}{*}{0.200} & AUC        & 1.000                     & 1.000                    & 0.996                     & 0.951                    & 0.999                    & 1.000                   & 1.000                    & 0.951                    \\
                         &                        & { T@1F} & 1.000                     & 0.998                    & 0.967                     & 0.802                    & 0.992                    & 1.000                   & 0.980                    & 0.430                    \\
                         & \multirow{2}{*}{0.300} & AUC        & 1.000                     & 1.000                    & 0.988                     & 0.939                    & 0.996                    & 1.000                   & 0.997                    & 0.917                    \\
                         &                        & { T@1F} & 1.000                     & 0.995                    & 0.890                     & 0.725                    & 0.964                    & 1.000                   & 0.960                    & 0.420                    \\
                         & \multirow{2}{*}{0.400} & AUC        & 0.999                     & 0.998                    & 0.976                     & 0.913                    & 0.982                    & 1.000                   & 0.999                    & 0.844                    \\
                         &                        & { T@1F} & 1.000                     & 0.989                    & 0.805                     & 0.634                    & 0.878                    & 0.997                   & 0.980                    & 0.150                    \\
                         & \multirow{2}{*}{0.500} & AUC        & 0.999                     & 0.998                    & 0.973                     & 0.901                    & 0.985                    & 1.000                   & 0.992                    & 0.832                    \\
                         &                        & { T@1F} & 1.000                     & 0.976                    & 0.759                     & 0.534                    & 0.879                    & 0.997                   & 0.840                    & 0.060                    \\
                         & \multirow{2}{*}{0.600} & AUC        & 0.999                     & 0.994                    & 0.952                     & 0.878                    & 0.979                    & 0.999                   & 0.982                    & 0.785                    \\
                         &                        & { T@1F} & 0.998                     & 0.958                    & 0.628                     & 0.474                    & 0.867                    & 0.993                   & 0.820                    & 0.090                    \\
                         & \multirow{2}{*}{0.700} & AUC        & 0.999                     & 0.992                    & 0.932                     & 0.851                    & 0.967                    & 0.998                   & 0.980                    & 0.771                    \\
                         &                        & { T@1F} & 0.998                     & 0.935                    & 0.556                     & 0.395                    & 0.741                    & 0.978                   & 0.810                    & 0.130                    \\
                         & \multirow{2}{*}{0.800} & AUC        & 0.997                     & 0.981                    & 0.891                     & 0.806                    & 0.946                    & 0.994                   & 0.966                    & 0.669                    \\
                         &                        & { T@1F} & 0.984                     & 0.833                    & 0.355                     & 0.338                    & 0.491                    & 0.885                   & 0.810                    & 0.120                    \\
                         & \multirow{2}{*}{0.900} & AUC        & 0.986                     & 0.927                    & 0.806                     & 0.762                    & 0.891                    & 0.964                   & 0.879                    & 0.540                    \\
                         &                        & { T@1F} & 0.880                     & 0.481                    & 0.203                     & 0.261                    & 0.348                    & 0.559                   & 0.190                    & 0.020                    \\ \hline
\end{tabular}
\caption{Robustness for LBW-Hard for VAR, VQ-GAN and RAR across different $\gamma$, ranging from 0.1 to 0.9.}
\label{hard}
\end{table*}

\begin{table*}[tbp] \small
\centering
\begin{tabular}{c|c|c|cccccccc}
\hline
                         & $\sigma$               & Metric     & Clean & Gaus  & Color & Geom  & JPEG  & VAE   & Diff  & Ctrl  \\ \hline
\multirow{18}{*}{VAR}    & \multirow{2}{*}{1.000} & AUC        & 0.927 & 0.852 & 0.873 & 0.505 & 0.844 & 0.745 & 0.632 & 0.543 \\
                         &                        & { T@1F} & 0.615 & 0.341 & 0.394 & 0.021 & 0.339 & 0.030 & 0.000 & 0.000 \\
                         & \multirow{2}{*}{2.000} & AUC        & 0.990 & 0.963 & 0.969 & 0.502 & 0.947 & 0.949 & 0.739 & 0.523 \\
                         &                        & { T@1F} & 0.940 & 0.764 & 0.833 & 0.022 & 0.751 & 0.450 & 0.120 & 0.000 \\
                         & \multirow{2}{*}{3.000} & AUC        & 0.997 & 0.984 & 0.986 & 0.507 & 0.977 & 0.979 & 0.846 & 0.575 \\
                         &                        & { T@1F} & 0.984 & 0.904 & 0.929 & 0.020 & 0.878 & 0.840 & 0.300 & 0.010 \\
                         & \multirow{2}{*}{4.000} & AUC        & 0.998 & 0.993 & 0.992 & 0.519 & 0.984 & 0.989 & 0.882 & 0.622 \\
                         &                        & { T@1F} & 0.993 & 0.945 & 0.953 & 0.027 & 0.915 & 0.870 & 0.400 & 0.000 \\
                         & \multirow{2}{*}{5.000} & AUC        & 0.999 & 0.995 & 0.993 & 0.650 & 0.988 & 0.992 & 0.844 & 0.620 \\
                         &                        & { T@1F} & 0.994 & 0.961 & 0.966 & 0.274 & 0.931 & 0.920 & 0.390 & 0.010 \\
                         & \multirow{2}{*}{6.000} & AUC        & 0.999 & 0.995 & 0.994 & 0.647 & 0.986 & 0.992 & 0.874 & 0.589 \\
                         &                        & { T@1F} & 0.995 & 0.961 & 0.969 & 0.274 & 0.933 & 0.920 & 0.470 & 0.010 \\
                         & \multirow{2}{*}{7.000} & AUC        & 1.000 & 0.995 & 0.994 & 0.665 & 0.989 & 0.995 & 0.892 & 0.626 \\
                         &                        & { T@1F} & 0.997 & 0.977 & 0.966 & 0.275 & 0.934 & 0.930 & 0.480 & 0.010 \\
                         & \multirow{2}{*}{8.000} & AUC        & 0.999 & 0.995 & 0.994 & 0.662 & 0.987 & 0.994 & 0.882 & 0.614 \\
                         &                        & { T@1F} & 0.995 & 0.966 & 0.968 & 0.283 & 0.933 & 0.920 & 0.460 & 0.000 \\
                         & \multirow{2}{*}{9.000} & AUC        & 0.999 & 0.994 & 0.993 & 0.665 & 0.986 & 0.990 & 0.879 & 0.617 \\
                         &                        & { T@1F} & 0.995 & 0.967 & 0.968 & 0.275 & 0.934 & 0.910 & 0.470 & 0.000 \\ \hline
\multirow{18}{*}{VQ-GAN} & \multirow{2}{*}{1.000} & AUC        & 0.986 & 0.959 & 0.786 & 0.786 & 0.901 & 0.990 & 0.993 & 0.669 \\
                         &                        & { T@1F} & 0.952 & 0.733 & 0.303 & 0.360 & 0.539 & 0.918 & 0.980 & 0.060 \\
                         & \multirow{2}{*}{2.000} & AUC        & 1.000 & 0.993 & 0.899 & 0.911 & 0.988 & 1.000 & 0.988 & 0.758 \\
                         &                        & { T@1F} & 0.998 & 0.969 & 0.620 & 0.576 & 0.921 & 1.000 & 0.990 & 0.180 \\
                         & \multirow{2}{*}{3.000} & AUC        & 0.999 & 0.990 & 0.925 & 0.956 & 0.991 & 0.998 & 0.985 & 0.829 \\
                         &                        & { T@1F} & 0.999 & 0.976 & 0.735 & 0.709 & 0.960 & 0.996 & 0.940 & 0.380 \\
                         & \multirow{2}{*}{4.000} & AUC        & 0.999 & 0.990 & 0.934 & 0.966 & 0.995 & 0.998 & 0.994 & 0.915 \\
                         &                        & { T@1F} & 0.998 & 0.977 & 0.770 & 0.776 & 0.977 & 0.998 & 0.990 & 0.610 \\
                         & \multirow{2}{*}{5.000} & AUC        & 0.999 & 0.988 & 0.924 & 0.962 & 0.993 & 0.998 & 0.997 & 0.887 \\
                         &                        & { T@1F} & 0.998 & 0.976 & 0.749 & 0.773 & 0.969 & 0.998 & 0.980 & 0.470 \\
                         & \multirow{2}{*}{6.000} & AUC        & 0.999 & 0.986 & 0.917 & 0.966 & 0.991 & 0.998 & 0.990 & 0.889 \\
                         &                        & { T@1F} & 0.997 & 0.970 & 0.734 & 0.768 & 0.958 & 0.996 & 0.970 & 0.460 \\
                         & \multirow{2}{*}{7.000} & AUC        & 0.999 & 0.986 & 0.910 & 0.967 & 0.990 & 0.998 & 0.995 & 0.889 \\
                         &                        & { T@1F} & 0.997 & 0.968 & 0.719 & 0.773 & 0.947 & 0.996 & 0.960 & 0.460 \\
                         & \multirow{2}{*}{8.000} & AUC        & 0.999 & 0.986 & 0.907 & 0.965 & 0.990 & 0.998 & 0.986 & 0.879 \\
                         &                        & { T@1F} & 0.997 & 0.970 & 0.720 & 0.761 & 0.948 & 0.996 & 0.940 & 0.360 \\
                         & \multirow{2}{*}{9.000} & AUC        & 0.999 & 0.986 & 0.907 & 0.963 & 0.989 & 0.998 & 0.987 & 0.876 \\
                         &                        & { T@1F} & 0.997 & 0.970 & 0.720 & 0.768 & 0.947 & 0.996 & 0.930 & 0.360 \\ \hline
\multirow{18}{*}{RAR}    & \multirow{2}{*}{1.000} & AUC        & 0.961 & 0.912 & 0.802 & 0.758 & 0.865 & 0.925 & 0.799 & 0.632 \\
                         &                        & { T@1F} & 0.633 & 0.314 & 0.169 & 0.254 & 0.292 & 0.522 & 0.170 & 0.010 \\
                         & \multirow{2}{*}{2.000} & AUC        & 0.998 & 0.993 & 0.946 & 0.879 & 0.982 & 0.996 & 0.968 & 0.790 \\
                         &                        & { T@1F} & 0.986 & 0.930 & 0.639 & 0.522 & 0.800 & 0.981 & 0.720 & 0.130 \\
                         & \multirow{2}{*}{3.000} & AUC        & 1.000 & 0.999 & 0.984 & 0.929 & 0.998 & 1.000 & 0.994 & 0.915 \\
                         &                        & { T@1F} & 0.999 & 0.988 & 0.860 & 0.722 & 0.973 & 0.996 & 0.970 & 0.430 \\
                         & \multirow{2}{*}{4.000} & AUC        & 1.000 & 1.000 & 0.994 & 0.947 & 0.998 & 1.000 & 1.000 & 0.958 \\
                         &                        & { T@1F} & 0.999 & 1.000 & 0.940 & 0.794 & 0.980 & 0.999 & 1.000 & 0.620 \\
                         & \multirow{2}{*}{5.000} & AUC        & 1.000 & 1.000 & 0.997 & 0.953 & 1.000 & 1.000 & 1.000 & 0.958 \\
                         &                        & { T@1F} & 1.000 & 1.000 & 0.968 & 0.823 & 0.995 & 1.000 & 1.000 & 0.620 \\
                         & \multirow{2}{*}{6.000} & AUC        & 1.000 & 1.000 & 0.997 & 0.958 & 1.000 & 1.000 & 1.000 & 0.978 \\
                         &                        & { T@1F} & 1.000 & 1.000 & 0.971 & 0.842 & 0.998 & 1.000 & 1.000 & 0.770 \\
                         & \multirow{2}{*}{7.000} & AUC        & 1.000 & 1.000 & 0.998 & 0.958 & 1.000 & 1.000 & 1.000 & 0.961 \\
                         &                        & { T@1F} & 1.000 & 1.000 & 0.978 & 0.851 & 0.997 & 1.000 & 1.000 & 0.740 \\
                         & \multirow{2}{*}{8.000} & AUC        & 1.000 & 1.000 & 0.999 & 0.961 & 1.000 & 1.000 & 1.000 & 0.978 \\
                         &                        & { T@1F} & 1.000 & 1.000 & 0.981 & 0.844 & 0.999 & 1.000 & 0.990 & 0.760 \\
                         & \multirow{2}{*}{9.000} & AUC        & 1.000 & 1.000 & 0.998 & 0.957 & 1.000 & 1.000 & 1.000 & 0.976 \\
                         &                        & { T@1F} & 1.000 & 1.000 & 0.980 & 0.847 & 0.997 & 1.000 & 1.000 & 0.770 \\ \hline
\end{tabular}
\caption{Robustness for LBW-Soft for VAR, VQ-GAN and RAR across different $\sigma$, ranging from 1 to 9.}
\label{soft}
\end{table*}

\section{More Visual Results}
% \textbf{Effect of $\gamma$ on Image Quality}

% \textbf{Efffect of $\sigma$ on Image Quality}

\textbf{Comparison between LBW-Post and other post-hoc methods.}

\begin{figure*}[thbp]
    \centering
    \includegraphics[width=0.9\textwidth]{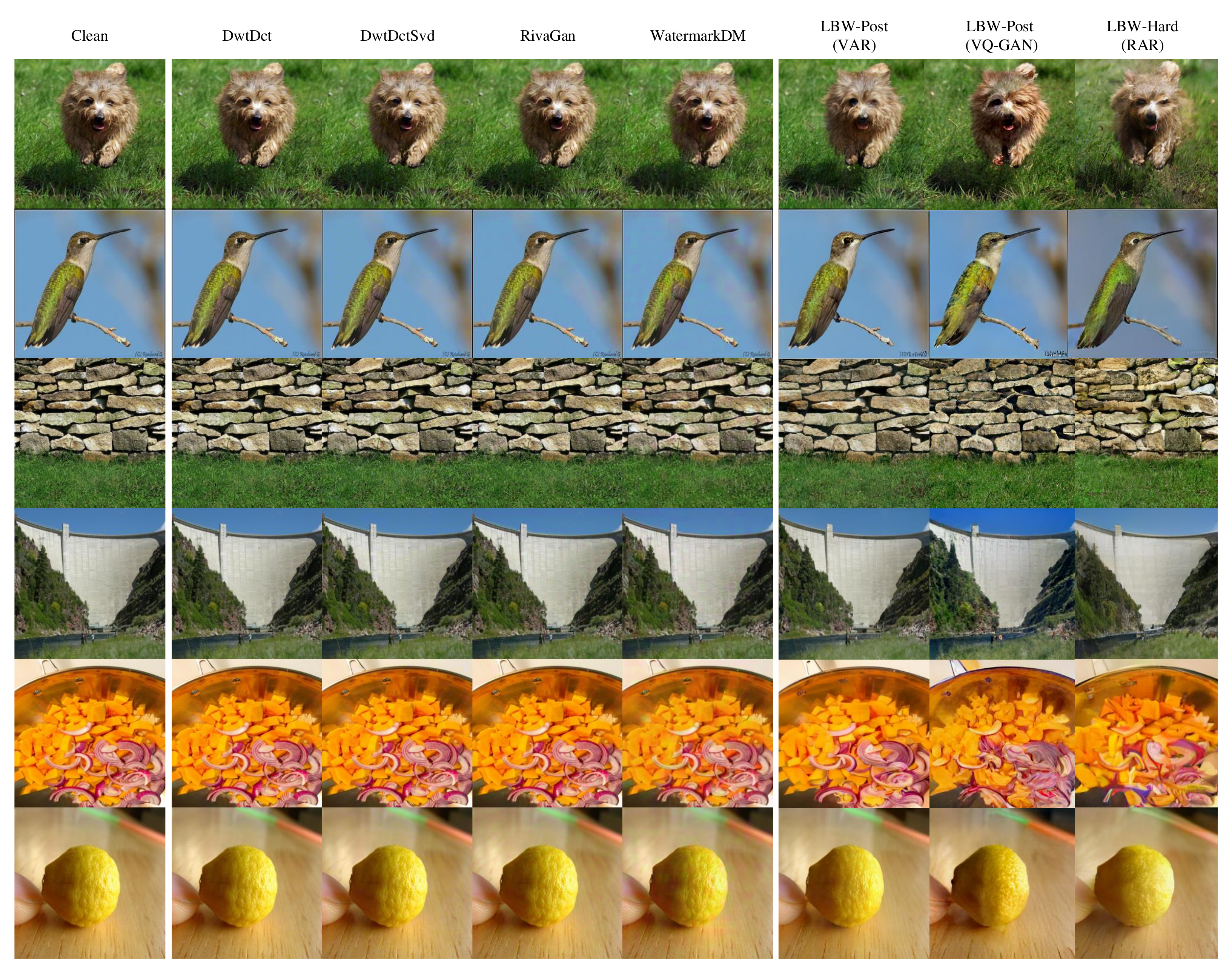}
    \caption{Visual comparison between our LBW-Post and other Post-hoc methods. }
    \label{fig:lbw_post_vs_posthoc}
\end{figure*}
Figure~\ref{fig:lbw_post_vs_posthoc} compares the watermarking performance of LBW-Post and traditional Post-hoc methods across different generative models (VQ-GAN, VAR, and RAR). In this comparison, LBW-Post employs a green word ratio of $\gamma=0.1$. Experimental results indicate that in multi-scale token map models (e.g., VAR), LBW-Post achieves image quality comparable to traditional Post-hoc methods. In single-scale generative models (e.g., VQ-GAN and RAR), LBW-Post also has a minimal impact on image quality, effectively maintaining visual consistency.

Furthermore, LBW-Post introduces an adjustable hyperparameter $\gamma$ (green word ratio) to regulate the trade-off between watermark embedding strength and image quality. This allows users to fine-tune the embedding strategy according to specific application requirements, balancing image quality and watermark robustness, thereby adapting to various practical scenarios.

\noindent\textbf{More Comparison between LBW-Hard and LBW-Soft.}
\begin{figure*}[hbp]
    \centering
    \includegraphics[width=0.9\textwidth]{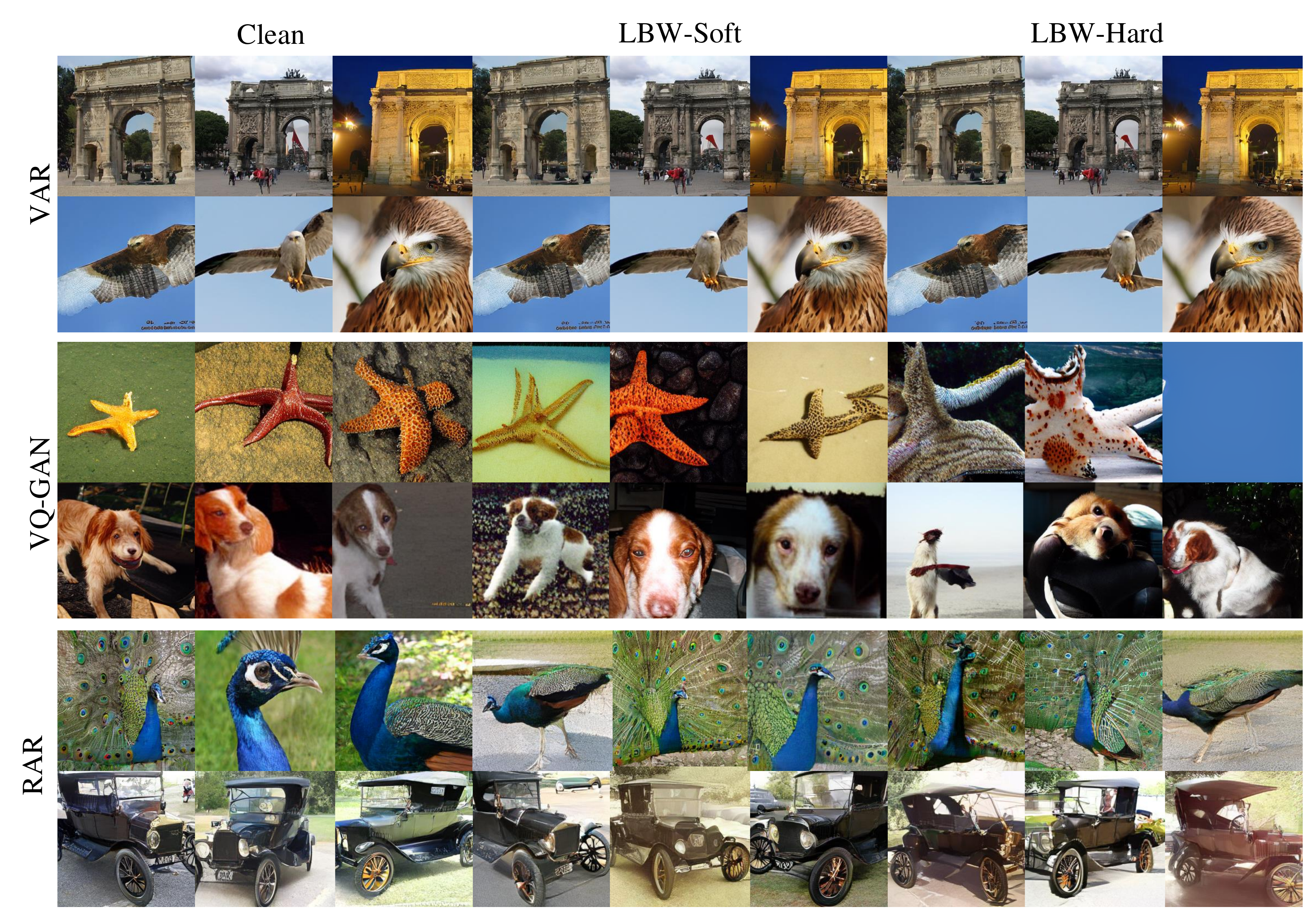}
    \caption{Comparison between LBW-Hard and LBW-Soft across VAR, VQ-GAN, and RAR.}
    \label{fig:lbw_hard_vs_soft}
\end{figure*}

Figure~\ref{fig:lbw_hard_vs_soft} compares the performance of LBW-Hard and LBW-Soft across different generative models. The results indicate that for AR models utilizing multi-scale token maps (e.g., VAR), both methods yield visually comparable results. However, for single-scale token map models (e.g., VQ-GAN and RAR), LBW-Soft outperforms LBW-Hard, producing images with richer details and stronger class relevance.

Moreover, LBW-Soft demonstrates superior robustness when the green word ratio is low, a scenario in which LBW-Hard may fail to generate meaningful images. This highlights LBW-Soft as a more adaptable solution that ensures stable image synthesis even under challenging conditions.


\end{document}